\documentclass[reprint,amsmath,amssymb,aip]{revtex4-1}
\usepackage{graphicx}
\usepackage{dcolumn}
\usepackage{bm}
\usepackage[usenames,dvipsnames]{color}
\usepackage{comment}
\newcommand \comma {\mbox{\makebox[.1 in]{ },}}

\def \be {\begin{equation}}
\def \ee {\end{equation}}
\def \ben {\begin{eqnarray}}
\def \een {\end{eqnarray}}
\def \bi {\begin{itemize}}
\def \ei {\end{itemize}}

\def\rc#1{{\color{black} {#1}}}
\usepackage{multirow}
\usepackage{ctable}
\begin{document}
\bibliographystyle{prsty}


\title{Perspective of Fermi's golden rule and its generalizations in chemical physics}

\author{Seogjoo J. Jang}
\email[ ]{seogjoo.jang@qc.cuny.edu}

\affiliation{Department of Chemistry and Biochemistry, Queens College, City University of New York, 65-30 Kissena Boulevard, Queens, New York 11367, USA \& PhD Programs in Chemistry and Physics, Graduate Center of the City University of New York, New York 10016, USA }

\author{Goun Kim}
\affiliation{Department of Chemistry, Korea Advanced Institute of Science and Technology (KAIST), Daejeon 34141, Korea}

\author{Young Min Rhee}
\affiliation{Department of Chemistry, Korea Advanced Institute of Science and Technology (KAIST), Daejeon 34141, Korea}
\date{Accepted for publication in {\it the Journal of Chemical Physics} on May 28, 2026}

\begin{abstract}
This perspective provides a succinct history of Fermi's golden rule (FGR), overview of its derivation, assumptions, and representative forms.  Major applications of FGR, mostly in the field of chemical physics, are reviewed.   These illustrate the broad applicability and success of FGR.  Ambiguities and open issues encountered in practical applications of FGR are clarified.   Recent advances in generalizations of FGR and computational methods for practical applications are addressed.    
\end{abstract}

\maketitle
\section{Introduction}
Fermi's golden rule (FGR)\cite{dirac-prsa114,schiff,fermi,heitler,cohen-t,sakurai-qm,schatz-ratner-qm-in-chem,jang-qmc} is the simplest but most widely used theory for calculating rates of quantum transitions. 
This name originated from Fermi's lecture on nuclear physics,\cite{fermi} where it was called\footnote{In Fermi's book, {``Golden Rule No. 1"}  denotes the rate expression now known as super-exchange rate.}  ``Golden Rule No. 2"  citing Schiff's book\cite{schiff} as a reference for its derivation.    However, the actual history of FGR is almost as old as that of quantum mechanics.  The first derivation of FGR can be traced back to Dirac's pioneering paper  on the emission and absorption of radiation,\cite{dirac-prsa114} \rc{a foundational work cited in earlier publications\cite{schiff,visser-ajp77,kramers,condon-shortley-1st} as the source of FGR} and  further extended by Fermi,\cite{fermi-rmp4} Heitler,\cite{heitler} and many others.  Thus, FGR  has been well recognized and  used from the early days of quantum mechanics,  helping elucidate mechanisms and establish selection rules in nuclear processes\cite{fermi} and  spectroscopies of atomic and molecular systems.\cite{condon-shortley-1st,breene,flygare,craig,steinfeld,cohen-t,schatz-ratner-qm-in-chem}  

\begin{figure*}
    \centering
    \includegraphics[width=\textwidth]{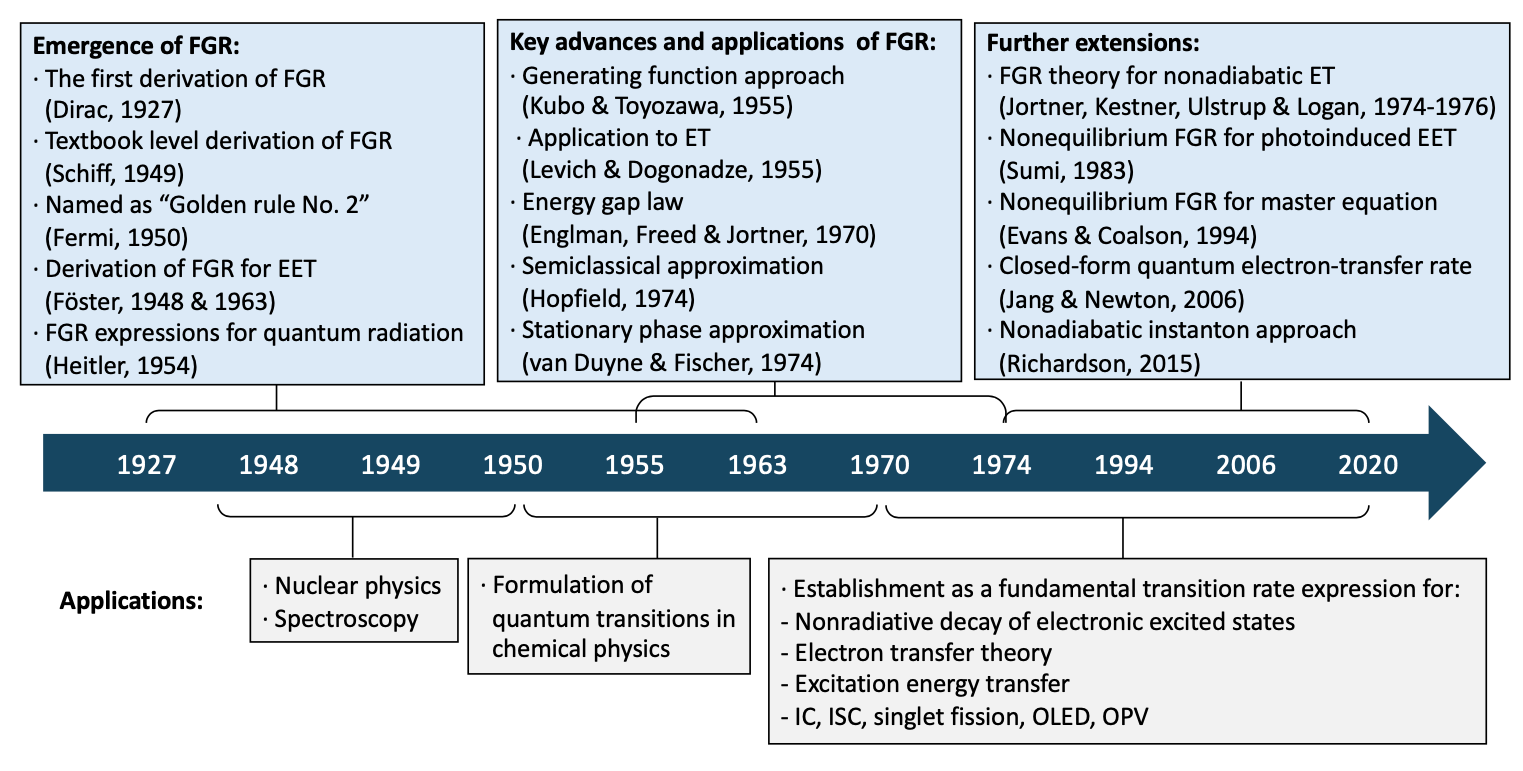}
    \caption{Some milestones in the development and applications of FGR.}
    \label{fig:timeline}
\end{figure*}

In 1940s and early 1950s, applications of FGR were  extended to other quantum transitions.\cite{goodman-pr71,forster-ap,kubo-pr86,kubo-ptp13,lax-jcp20,dexter-jcp} Most of these pioneering works however referred to FGR simply as an obvious outcome of the time dependent perturbation theory\cite{schiff,heitler,cohen-t,sakurai-qm,schatz-ratner-qm-in-chem} without offering detailed explanation. An exception to this was Förster's work,\cite{forster-ap,forster-book} which in itself serves as a detailed derivation of FGR (and the time dependent perturbation theory) for electronic energy transfer (EET) processes in molecular systems.  Applications of FGR further broadened in late 1950s and 1960s. Formative works in chemical physics during this period include the first quantum formulation of electron transfer (ET) theory\cite{levich-dan124} and  the theory of non-radiative decay of electronic excited states.\cite{lin-jcp44,robinson-jcp37,bixon-jcp48}   Since 1970s, FGR and  its name seemed well established in the chemical physics community, and has been used widely for nonradiative decay of electronic excited states,\cite{englman-mp18,freed-jcp52,gelbart-jcp52,freed-acr11,nitzan-jcp56,lin-jcp58,fong-jcp56,nitzan-jcp63,lin-jcp58,jang-jcp155-1} ET theory,\cite{marcus-arpc15,kestner-jpc78,ulstrup-jcp63,jortner-jcp64,marcus-bba811,siders-jacs103,siders-jacs103-2,marcus-jpc86,marcus-jcp81,schmicker-ea18,onuchic-jpc90,adv-et,newton-cr91,nitzan,medvedev-jcp107,jang-jcp122,jang-jpcb110,basilevsky-jcp139} EET\cite{silbey-arpc27,may,jang-cp275,jang-prl92,jang-jcp127,jang-wires3,jang-exciton} and various other processes. 
See Fig.\ \ref{fig:timeline} for a schematic view on this development history.

Despite the status of FGR as a well-defined textbook-level theory and the history of its numerous applications, there remain subtle issues\cite{jang-jcp159,jang-jcp161} and computational challenges in practical applications of  FGR.   
There have also been further extensions\cite{coalson-jcp101,evans-jcp104,jang-cp275,jang-jcp122,jang-jcp127,jang-prl92,jang-jpcb110,jang-wires3,jang-prl113,sun-jctc12,sun-jcp144,sun-jcp145,jang-jcp159,jang-jctc21} of FGR, making applications of the underlying time dependent perturbation theory much more effective than was originally perceived.   Considering these, it is timely to revisit assumptions involved in FGR, to assess its success and unsettled issues,  and to offer a  review of its recent generalizations.  These are objectives of this perspective.  

We here provide a brief overview of FGR and its underlying assumptions, followed by a review of its key applications.   We will also go over some select set of theories allowing further generalizations of FGR beyond its intended regime. Experimental studies will also be provided to illustrate the scope of the applicability of FGR.  
 \rc{The issues and topics we address here are guided by our experience and interest in the field of chemical physics.  Thus, other important areas beyond our expertise are largely omitted, which include quantum scattering,\cite{chattopadhyay-rmp53,marchetti-jpcm30} sticking,\cite{clougherty-prb46,zhang-prl108,clougherty-prb112} transport,\cite{koller-prb82,ferguson-prr3} and many-body theories.\cite{micklitz-prl129}} 

\rc{The paper is organized as follows.  Section II reviews the basic derivation of FGR, followed by summaries of underlying assumptions and subtle issues.  Section III provides an overview of different expressions for FGR and their implications and applications.  Section IV provides further approximations of FGR that are important for practical applications. Section V provides applications and generalizations of FGR.  Section VI addresses environmental effects, and Sec. VII concludes the paper with a few concluding remarks.}   
 
\section{Basic derivation, assumptions, and ambiguities\label{section2}}     
Although well established, it is instructive to provide an overview of the derivation of FGR and its underlying assumptions.  Given  an initial quantum \rc{state} $|\psi_j\rangle$ and a final state $|\psi_f\rangle$, which are well-defined eigenstates of a zeroth order Hamiltonian $\hat H_0$ and are coupled only through a perturbation Hamiltonian $\hat H_c$, the elementary FGR rate for the quantum transition between the two states is
\be
k_{_{{\rm F},j\rightarrow f}}=\frac{2\pi}{\hbar}\left |\langle \psi_f|\hat H_c|\psi_j\rangle \right |^2 \delta (E_j-E_f),  \label{eq:kfgr-0}
\ee
where $E_j$ and $E_f$ are eigenvalues of $|\psi_j\rangle$ and $|\psi_f\rangle$. 
The Dirac-delta function $\delta (E_j-E_f)$ on the righthand side of the above equation is understood as the following Fourier representation:
\rc{\be
\delta (E_j-E_f)=\lim_{t_s\rightarrow \infty} \frac{1}{2\pi\hbar}\int_{-t_s}^{t_s} dt\ e^{i(E_j-E_f)t/\hbar} . \label{eq:delta}
\ee}

Appendix \ref{app-a} provides a detailed derivation of  Eq. (\ref{eq:kfgr-0}), and clarifies that it implicitly assumes the existence of  a long enough stationary time $t_s$ satisfying the following condition:
\begin{enumerate} 
\item [{\bf C1}.]The first order approximation for the transition amplitude, 
\be
c_{jf}^{(1)}=-\frac{i}{\hbar}\int_0^{t_s} dt \langle \psi_f| e^{-i\hat H_0 (t_s-t)/\hbar}\hat H_c e^{-i\hat H_0 t/\hbar}|\psi_j\rangle , \label{eq:tr_am-1st}
\ee 
remains accurate enough during $t_s$.
\end{enumerate}
It is also important to note that the Dirac-delta function in Eq. (\ref{eq:kfgr-0}) is meaningful only under the assumption that the final states form a {\it continuum of states} with energy density $\rho _f(E)$.  Furthermore, application of FGR is possible only if the following condition is met.  
 \begin{enumerate}
\item [{\bf C2}.] The relaxation among the final states with energy density $\rho _f(E)$ occurs much faster than $t_s$.  
\end{enumerate} 
With this assumption,  $\delta (E_j-E_f)$ in Eq. (\ref{eq:kfgr-0}) can be replaced with $\rho_f(E_j)$.   Additional averaging of the resulting expression over the distribution of initial states, with probability $p_j$ for each  $|\psi_j\rangle$, leads to the following averaged FGR rate expression:
\be
k_{_{\rm F}}=\frac{2\pi}{\hbar}\sum_j p_j \left |\langle \psi_f|\hat H_c|\psi_j\rangle \right |^2 \rho_f (E_j) .  \label{eq:kfgr-g}
\ee
Note that this expression can still be viewed as the most general one since Eq. (\ref{eq:kfgr-0}) corresponds to  a special case where there is a single initial state $j$ and $\rho_f(E_j)$ becomes a delta function. 

The FGR rate expression, Eq. (\ref{eq:kfgr-g}),  reduces the task of calculating a key quantum dynamical quantity to simpler static and statistical calculations.    Namely, FGR provides an answer for the rate of quantum dynamical change only using the information on the matrix element $\langle \psi_f|\hat H_c|\psi_j\rangle$ of the coupling Hamiltonian and the distribution of initial and final states.  However, even with this simplification, challenges remain in practical applications of FGR for molecular systems in condensed or complex environments due to the following four factors:
\begin{enumerate}
\item [{\bf F1.}] Exact information on the true zeroth order Hamiltonian $\hat H_0$, which in principle should include all the environmental effects as well as the primary system of interest, is not fully known or difficult to consider exactly.   
\item [{\bf F2.}] Identification of the proper coupling Hamiltonian $\hat H_c$ and calculation of its matrix elements, with respect to all coupled initial and final states, are challenging  in general.
\item [{\bf F3.}] Correct and efficient determination of the initial distribution of states and the energy density of final states $\rho_f(E)$ is challenging in general, especially when marginally small statistical events play important roles and/or because the full source determining $\rho_f(E)$ is generally unknown in actual applications.
\item [{\bf F4.}] The assumption of the existence of $t_s$ and the validity of assuming a rate behavior are not easy to justify {\it a priori} in general, and are often difficult to assess quantitatively even for relatively simple model systems unless exact quantum dynamics calculations are available.      
\end{enumerate}
Although listed separately, the above four factors are not necessarily independent but are often interrelated with each other in different ways depending on the nature of quantum processes. Thus, addressing these factors and formulating more refined rate expressions that can be determined more easily, often with additional approximations, have been central tasks for developing different rate theories dealing with different processes.  

There are also some ambiguities and subtle issues that arise in applications and interpretations of FGR.  In relation to the factors {\bf F1} and {\bf F3} above, the proper choice of  $\rho_f(E_j)$ is debated quite often in practical applications.   This remains true despite recent advances in more accurate evaluation of FGR rates for more complex molecular systems.\cite{peng-jcp126,etinski-jcp134,niu-scc51,niu-jpca114,baiardi-jcp144,kim-jctc16,wang-jcp154} \rc{A recent work\cite{debierre-prsa477} also demonstrated the relationship between the factors {\bf F3} and {\bf F4}, by analyzing the dependence of $t_s$ on $\rho_f(E)$ for simple model cases. } Another important point to recognize is that FGR  is typically used in the context of master equation for population dynamics, which is not considered in standard textbook level derivations, often with much better accuracy than anticipated from a simple analysis of the perturbation theory.  There are multitude of reasons for this, but the contribution of decoherence due to couplings to molecular  and other environmental degrees of freedom seems to be the primary one in many cases.    As yet, proving this is challenging because reliable calculation of full decoherence effects is difficult due to the lack of exact results beyond simple model Hamiltonians.

\section{Alternative expressions for FGR}
\rc{Although the energy domain expression, Eq. (\ref{eq:kfgr-g}), contains all the information necessary for the evaluation of the FGR, its practical evaluation is limited to simple model systems or few dimensional molecular systems.  This is because of the numerical difficulty of calculating eigenstates and eigenvalues and enumerating all the eigenstates. Alternative expressions have been developed to overcome or bypass these issues. }
\subsection{Time-domain expression of FGR}
With the advance of quantum dynamics methods, direct calculation of time domain expression for FGR rate has become feasible and popular.  The time domain expression is often derived by combining the  standard energy domain expression Eq. (\ref{eq:kfgr-0}) with Eq. (\ref{eq:delta}), but is  in fact  a precursor of Eq. (\ref{eq:kfgr-0}) as detailed in Appendix \ref{app-a}.  
Let us introduce the density operator for the final state, $\hat \rho_f=|\psi_f\rangle\langle \psi_f|$.  Then, Eq. (\ref{eq:kif-time}) can be expressed as 
\ben
 &&k_{_{j\rightarrow f}}(t_s) \nonumber \\
 &&=\frac{2}{\hbar^2} {\rm Re} \int_0^{t_s} dt\ {\rm Tr} \left\{ \hat H_c^\dagger (t_s) \hat \rho_f e^{-i\hat H_0 t/\hbar}\hat H_c(t_s-t) \hat \rho_j e^{i\hat H_0 t/\hbar}  \right\} , \nonumber \\ \label{eq:kif-time-1}
 \een
 where $\hat \rho_j=|\psi_j\rangle\langle \psi_j|$ as has been introduced in App. \ref{app-a}.

For time independent $\hat H_c$  and in the  \rc{limit} of $t_s\rightarrow \infty$, the above expression reduces to the following time-domain expression for FGR: 
\ben
 &&k_{_{{ \rm F},j\rightarrow f}} \nonumber \\
 &&=\frac{2}{\hbar^2} {\rm Re} \int_0^{\infty} dt\ {\rm Tr} \left\{ \hat H_c^\dagger \hat \rho_f e^{-i\hat H_0 t/\hbar}\hat H_c \hat \rho_j e^{i\hat H_0 t/\hbar}  \right\} . \label{eq:kfgr-time-r}
 \een
 
For the case where $\hat H_c(t)=\hat h_ce^{-i\omega t}+\cdots$ and \rc{the contribution of terms in $\cdots$ is negligible (rotating wave approximation)}, Eq. (\ref{eq:kif-time-1}) in the limit of $t_s \rightarrow \infty$ reduces to the following time-domain expression:
\ben
 &&k_{_{{\rm F}, j\rightarrow f}}(\omega) \nonumber \\
 &&=\frac{2}{\hbar^2} {\rm Re} \int_0^{\infty} dt\  e^{i\omega t}{\rm Tr} \left\{ \hat h_c^\dagger \hat \rho_f e^{-i\hat H_0 t/\hbar}\hat h_c \hat \rho_j  e^{i\hat H_0 t/\hbar} \right\} . \label{eq:kfgr-time-s}
 \een
 This is a key expression used for the lineshape calculation \rc{of electronic processes}, as will be explained in more detail later.
 
\rc{For coupled harmonic oscillator systems, well established closed form expressions\cite{lin-jcp44,lin-jcp58,yan-jcp85,mebel-jpca103,ianconescu-jpca108,niu-jpca114,peng-cp370,borrelli-jpca116,Banerjee2017jctc,miyazaki-jcp156} are available for integrands of Eqs. (\ref{eq:kfgr-time-r}) and (\ref{eq:kfgr-time-s}), which have long played central roles in modeling lineshapes and rates.  For more general systems with anharmonic potentials, direct time evolution methods can be used.\cite{neria-prl67,neria-cp183,shi-jpc108,sun-jpca120,saller-jpcl13,makri-arpc50,baltaretu-jcp133,huo-jcp137,makri-jcp148,beck-pr324,meyer-wires2,wang-jcp146}  Although most of these quantum dynamics methods are approximate by design or in practice, they can be improved in a systematic and controllable manner, resulting in fairly accurate results. Furthermore, if the precursor of FGR, Eq. (\ref{eq:kif-time-1}), is used instead, these dynamics methods  can easily be extended to calculate nonequilibrium generalizations of the FGR rate directly.  }

\rc{The benefit of the time domain expression, as noted above, comes from the fact that predetermination of eigenvalues and eigenstates is not necessary.   However, it is noteworthy to point out that identifying good approximations for $\hat \rho_j$ and $\hat \rho_f$ can still be critical in ensuring the accuracy of calculated rates.  Practical and successful methods of approximation have been developed to this end.  These include approximations with Gaussian wave packets, representation with appropriate Wigner distributions, and sampling with imaginary time path integral simulation methods.}  

\rc{An  important practical issue that requires careful assessment in direct numerical calculation of time domain expressions is the error due to truncation of time integration at a finite upper limit,\cite{miyazaki-jcp156} which is often inevitable, or the lack of convergence due to persistent oscillatory terms.  This latter issue becomes prevalent  for finite dimensional systems of coupled harmonic oscillators and has been typically handled by phenomenological approaches.  Further discussion of this issue is provided in the next subsection.   } 

\subsection{Generating function approach for FGR}
\begin{figure}
    \centering
    \includegraphics[width=\linewidth]{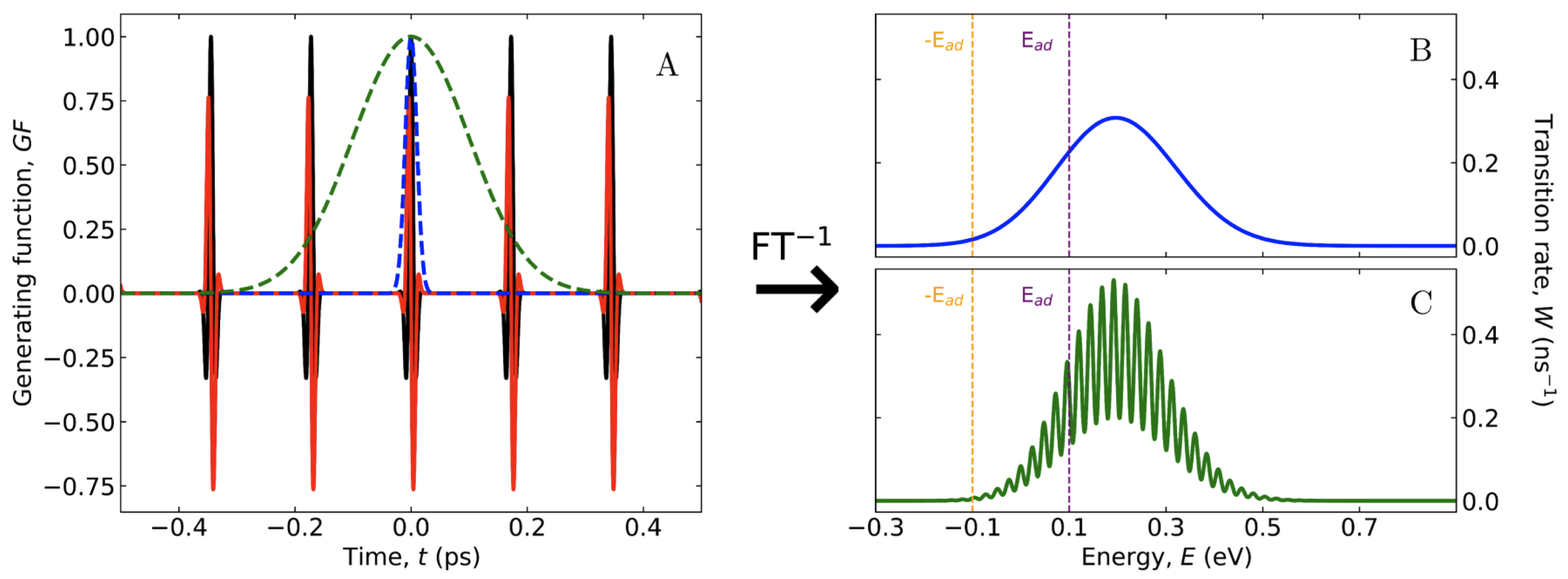}
    \caption{(A) The real (black) and imaginary (red) parts of the time-dependent GF in Eq.\ (\ref{eq:kF_GF}) for a system with single vibrational mode and two diabatic electronic states. The dashed lines indicate Gaussian apodization functions with two different widths. (B) The rate spectrum with narrower apodization function, and (C) the rate spectrum with the wider one. In the rate spectra, vertical dashed lines represent the locations of adiabatic gaps for upward and downward transitions, with the gap denoted as $E_{ad}$. Reproduced with permission from J. Phys. Chem. C {\bf 128}, 18598 (2024) [Ref. \onlinecite{landi-jpcc128}]. Copyright  2024  American Chemical Society. }
    \label{fig:GF}
\end{figure}

Kubo and Toyozawa (KT)\cite{kubo-ptp13} developed a generating function (GF) approach for the FGR rate, which has been widely adopted, by extending an earlier work by Kubo.\cite{kubo-pr86}  To this end, KT first defined\cite{kubo-ptp13} the following energy dependent rate function:
\be
k_{_{\rm F}} (E)= \frac{2\pi}{\hbar} \sum_j \sum_f  \frac{e^{-\beta E_j}}{Z_i(\beta)} |\langle \psi_f|\hat H_c|\psi_j\rangle|^2 \delta (E-E_f+E_j) ,
\ee 
where $Z_i(\beta)=\sum_j e^{-\beta E_j}$.  Note that $k_{_{\rm F}} (E=\hbar\omega)$ is the spectroscopic transition rate and $k_{_{\rm F}} (0)$ is the conventional FGR (without external perturbation) for the equilibrium canonical distribution as the initial distribution. Then, they introduced the (moment) generating function for $k_{_{\rm F}} (E)$  as the following two-sided Laplace transform:\cite{kubo-ptp13}
\ben
f(\lambda)&=&\int_{-\infty}^{\infty} k_{_{\rm F}} (E) e^{-\lambda E} dE \nonumber \\
&=&\frac{2\pi}{\hbar} \sum_j \sum_f  \frac{e^{-(\beta -\lambda)E_j}}{Z_i(\beta)} |\langle \psi_f|\hat H_c|\psi_j\rangle|^2 e^{-\lambda E_f} \nonumber \\
&=&\frac{1}{Z_i (\beta)}{\rm Tr} \left\{\hat H_c^\dagger e^{-\lambda \hat H_0}\hat H_c e^{-(\beta-\lambda)\hat H_0} \right \}.  \label{eq:gen-ke}
\een
Note that this is nothing but the integrand of Eq. (\ref{eq:kfgr-time-r}) with $t=-i\hbar \lambda$ for the case where $\hat \rho_j$  is a canonical density operator of the zeroth order Hamiltonian $\hat H_0$ (projected on to the subspace of initial states) and $\rho_f$ is an identity operator.  This equivalence can be understood from the fact that the double-side Laplace transform and the Fourier transform are related via an analytic continuation. 
In other words, $k_{_{\rm F}} (E)$ is related to $f(-it/\hbar)$ as follows:
\be
k_{_{\rm F}} (E)=\frac{1}{\hbar^2}\int_{-\infty}^\infty dt e^{iEt/\hbar} f(-it/\hbar) . \label{eq:kF_GF}
\ee
\rc{Most applications\cite{freed-jcp52,freed-acr11,Banerjee2017jctc}  of the GF approach that involve numerical evaluation have used the above expression} due to the convenience of Fourier transform, 
which is equivalent to calculating  the integrand of the time domain expression for the FGR \rc{as described in Sec. IIIB}.  In fact, this has been well recognized by many experts early on although the view that the GF approach offers a unique numerical advantage persists.    This view has its point since the ``GF-based numerical approaches" in general intend to account for all the effects of molecular vibrations explicitly in calculating $f(-it/\hbar)$. \rc{For this, closed form expressions for coupled multidimensional harmonic oscillators have been actively utilized.\cite{kubo-ptp13,sharp-jcp41,lin-jcp44,lin-jcp58,yan-jcp85,mebel-jpca103,ianconescu-jpca108,niu-jpca114,peng-cp370,borrelli-jpca116,Banerjee2017jctc,miyazaki-jcp156}  However, as mentioned at the end of Sec. IIIB, for most finite and closed systems of coupled harmonic oscillators, oscillatory behavior of $f(-it/\hbar)$ makes it difficult to obtain a convergent result without introducing window or apodization functions.\cite{landi-jpcc128} This amounts to accounting for the effects of open environments in an empirical manner, a process that introduces ambiguity in determining rates.}      See Fig.\ \ref{fig:GF} for a representative example showing the effect of apodization.\cite{landi-jpcc128} \rc{A rationale for this procedure comes from the convolution expression described in the next subsection.}

\rc{\subsection{Convolution expressions}
When the integrand of the time domain expression or the generating function for FGR can be expressed as a product of two or more independent components,  the  lineshape or rate can be expressed as convolution expressions in the frequency domain.  In general, this serves as the basis for using smoothening or apodization functions given that there are additional (infinite) degrees freedom coupled weakly enough so as not to affect the dynamics of the system, while still contributing to the energy conservation requirement. }

\rc{Historically, convolution expressions have also played important roles in the development of key theories. F\"{o}rster's spectral overlap expression\cite{forster-ap,forster-book} for EET results from a convolution expression relying on the fact that the response of molecular vibrations and environments to the excitation of the energy donor are independent of those for the energy acceptor.   Kestner {\it et al.}\cite{kestner-jpc78} also employed the convolution expression to develop a general formulation of ET that incorporates the effects of quantum degrees of freedom and classical dielectric medium.  More recently, Cui {\it et al.}\cite{cui-jcp155} applied the convolution expression in developing a theory of heat transport accompanying electron transfer processes.  }

\subsection{Reduced expression for FGR for a system coupled to a bath}
In many applications, it is convenient to express the rate as a partial average over the uninteresting degrees of freedom, which is typically referred to as bath. 
For a generic case as detailed in App.  \ref{app-b}, one obtains Eq. (\ref{eq:kt-fi}) as a general time dependent rate.   For the case of time independent $\hat H_c$ (and thus time independent $\hat J_{jk}$ in Eq. (\ref{eq:hct-gen})), the FGR rate from a system state with energy ${\mathcal E}_j$ to another system state with energy ${\mathcal E}_f$ is obtained as the $t_s\rightarrow \infty$ limit of Eq. (\ref{eq:kt-fi}) and becomes
\ben
&&k_{_{{\rm F},j\rightarrow f}}=\frac{2}{\hbar^2} {\rm Re} \int_0^{\infty} d t e^{i({\mathcal E}_j-{\mathcal E}_f)t/\hbar} \nonumber \\
&&\hspace{.2in}\times   {\rm Tr}_b\left\{ \hat J_{jf}^\dagger  e^{-i (\hat B_f+\hat H_b) t/\hbar}\hat J_{jf}  \hat \rho_{b,j} e^{i t ( \hat B_j+\hat H_b)/\hbar } \right\} ,  \label{eq:kfgr-b-fi}
\een 
where ${\rm Tr}_b$ is the trace over the bath degrees of freedom, $\hat J_{jf}$ is a coupling between the initial and final system states, which depends on the bath degrees of freedom in general, $\hat B_j$ ($\hat B_f$) is the bath operator coupled to the system state $j$ ($f$), and $\hat H_b$ is the bath Hamiltonian.  More detailed description and definitions of these terms are provided in App. \ref{app-b}. 
For the spectroscopic transition where $\hat J_{jf}(t) =\hat J_{jf} e^{-i\omega t}$, the above expression remains the same except that $\hbar \omega$ is added to ${\mathcal E}_j-{\mathcal E}_f$.  
 
For the simple case where (i) the bath can be modeled as a set of linearly coupled harmonic oscillators and (ii)  $\hat J_{jf}$ is independent of the bath (so called  Condon approximation), Eq. (\ref{eq:kfgr-b-fi}) can be simplified further to the following expression:
 \be
k_{_{{\rm F},j\rightarrow f}}=\frac{2|J_{jf}|^2}{\hbar^2} {\rm Re} \left [ \int_0^\infty d t e^{i( \tilde {\mathcal E}_j- \tilde {\mathcal E}_f ) t/\hbar - {\mathcal C}_{jf}(t) } \right ]  , \label{eq:kf-3}
\ee
where $\tilde {\mathcal E}_j$ ($\tilde {\mathcal E}_f$) defined by Eq. (\ref{eq:rel-eng})  is the  relaxed energy of the system state $j$ ($f$)  due to the effect of the bath,  and ${\mathcal C}_{jf}(t)$ is the bath correlation function defined by Eq. (\ref{eq:ct}).  

Equation (\ref{eq:kf-3}) and its spectroscopic version (with $\pm \hbar\omega$ added to the energy difference) have played an important role for practical modeling of  rates and spectral line shapes for systems in the condensed phase. For the most widely used Ohmic model (with appropriate high frequency cutoff) for the bath spectral density defined by Eq. (\ref{eq:bath-sp}), the numerical value of Eq. (\ref{eq:kf-3}) is finite for any value of the energy difference.    However, for a super-Ohmic bath spectral density, let alone discrete bath spectral densities in general, Eq. (\ref{eq:ct}) still contains delta-function singularity, which makes it impossible to define a rate.  A modification\cite{jang-jcp159,jang-jcp161} handling this issue has been proposed recently, which requires Master equation (ME) level description as detailed in the next subsection.

\subsection{FGR as transition probabilities of MEs\label{me-fgr}}
Many quantum processes can be described in terms of the following ME for the time dependent population $p_j(t)$:
\ben
\frac{d}{dt} p_j (t)=\sum_{k\neq j} \big \{ p_k(t) {\mathcal W}_{k\rightarrow j}(t)  -  p_j (t) {\mathcal  W}_{j\rightarrow k}(t)\big \} , \label{eq:me}
\een 
where ${\mathcal W}_{j\rightarrow k}(t)$ is the time dependent transition rate from state $j$ to $k$. In general, the state index in this ME represents that of a particular system of interest, which is coupled to other degrees of freedom.   \rc{Appendix \ref{app-c} provides a complete derivation of the above ME starting from a general quantum Liouville equation, employing a projection operator formalism, and provides exact formal expressions} for ${\mathcal W}_{j\rightarrow k}(t)$. \rc{More comprehensive formulation\cite{jang-exciton} and alternative derivations focused on different aspects that go beyond FGR are available as well.\cite{trushechkin-jcp151,lai-jcp155,wilson-jcp163}} 

\rc{In practice,} approximations are needed for the evaluation of ${\mathcal W}_{j\rightarrow k}(t)$.  
As detailed in App. \ref{app-c}, Eq. (\ref{eq:kif-time-1}) is equivalent to ${\mathcal W}_{j\rightarrow k}^{(2)}(t)$ given by Eq. (\ref{eq:wjk-2}) under the same physical assumption.  
Thus, using the FGR expression for the transition rate amounts to making an additional assumption of time scale separation such that \rc{$|p_j(t+t_s)-p_j(t)| \ll p_j(t)$}. 
This assumption can be valid even beyond limits where the first order time dependent perturbation theory is valid,  which explains why the use of FGR within the above ME often gives a reasonable description of the dynamics when perturbation theory is not expected to work.  

Further generalizations of FGR including nonequilibrium effects and higher order corrections within the ME approach have been developed.\cite{sparpaglione-jcp88,evans-jcp104,jang-prl113,jang-exciton,golosov-jcp115-1,golosov-jcp115-2,golosov-cp296,wu-jcp139}  The definition of FGR in the context of the above ME also makes it possible to defined a modified FGR rate expression\cite{jang-jcp159,jang-jcp161} for the case where the original FGR rate is ill-defined because of the presence of time dependent fluctuations or the sparsity of the final density of state such as the super-Ohmic bath spectral density.

\section{Further approximations of FGR}
\subsection{Semiclassical and stationary phase approximations of FGR for bath of harmonic oscillators}
The term ``semiclassical approximation" (SCA) has had varied meaning since it can literally imply any approximation that accounts for some quantum aspects in a classical manner.   \rc{In particular, SCAs of quantum propagators have generated a broad class of real time quantum dynamics methods\cite{miller-jpca105,stock-prl78,bennun-jpca104,cotton-jctc12,liang-jcp149,provazza-jctc14,shi-jcp120} that can be combined with the time domain expression of FGR  in Sec. IIIB.  The term SCA in this subsection refers to earlier and more limited use as a simple closed-form approximation of FGR that accounts for minimal quantum effects.  For more advanced and refined definitions and treatments, readers can refer to recent works by Sun and Geva.\cite{sun-jctc12,sun-jpca120}}  

In the context of earlier approximation for the FGR rate expression for electron transfer (ET) theory,\cite{marcus-jcp81,marcus-bba811} SCA was mostly understood as using a quantum thermal broadening factor.\cite{kubo-ptp13,lax-jcp20,hopfield-pnas71,siders-jacs103,siders-jacs103-2}  
This SCA expression can be easily derived by considering the short time quadratic expansion of the exponent in Eq. (\ref{eq:kf-3}).  Introducing  $\Delta G_{jf}=\tilde E_f-\tilde E_j$, this results in
 \be
k_{_{{\rm F},j\rightarrow f}}\approx\frac{2|J_{jf}|^2}{\hbar^2} {\rm Re} \left [ \int_0^\infty d t e^{i \Delta G_{jf} t/\hbar - \lambda_{qc}t^2/(\beta\hbar^2) -i\lambda t/\hbar } \right ]  , \label{eq:kf-2-sca}
\ee
where the reorganization energy $\lambda$ and  a quantum reorganization energy $\lambda_{qc}$, which employs the ${\rm coth}$ function for quantum factor,\cite{hopfield-pnas71}  are respectively defined as  
\ben 
&&\lambda\equiv\hbar\sum_n \delta g_{n,jf}^2 \omega_n =\frac{1}{\pi}\int_0^\infty d\omega \frac{{\mathcal J}_{jf}(\omega)}{\omega} , \label{eq:lambda-def} \\
&&\lambda_{qc}\equiv \frac{\beta\hbar^2}{2}\sum_n \delta g_{n,jf}^2 \coth(\frac{\beta\hbar\omega_n}{2}) \omega_n^2 \nonumber \\
&&\hspace{.3in}=\frac{\beta\hbar}{2\pi}\int_0^\infty d\omega {\mathcal J}_{jf}(\omega) \coth(\frac{\beta\hbar\omega}{2}) .
\een
Note that both $\lambda$ and $\lambda_{qc}$ depend on the indices of initial and final states, $j$ and $f$, which however are not shown explicitly to simplify the notation. 

Integration of Eq. (\ref{eq:kf-2-sca}) results in the following SCA:\cite{hopfield-pnas71}    
\be
k_{_{{\rm F},j\rightarrow f}}^{\rm SCA}=|J_{jf}|^2 \sqrt{\frac{\pi\beta}{\hbar^2\lambda_{qc}}}  \exp\left \{ -\beta \frac{(\Delta G_{jf}+\lambda)^2}{4\lambda_{qc}}  \right\}  . \label{eq:kf-2-sca-1}
\ee
In the classical or high temperature limit where  \rc{$\beta\hbar\omega_n \ll 1$} for all $\omega_n$,  $\lambda_{qc}$ approaches $\lambda$ and the above expression reduces to the celebrated Marcus rate expression\cite{marcus-jcp43} given below:
\be
k_{_{M},j\rightarrow f}=|J_{jf}|^2 \sqrt{\frac{\pi\beta}{\hbar^2\lambda}}  \exp\left \{ -\beta \frac{(\Delta G_{jf}+\lambda)^2}{4\lambda }  \right\}  . \label{eq:kf-m}
\ee
Although this expression was used more broadly in the context of ET, it can also be applied to the EET process with a proper definition of reorganization energy.\cite{jang-cp275} Nonequilibrium extensions of this expression for both ET and EET have also been developed for photoinduced ET\cite{cho-jcp103} and for nonequilibrium RET theory for pump-probe spectroscopic setup.\cite{sumi-prl50,jang-cp275,jang-exciton}  
 
  \begin{figure}
    \centering
    \includegraphics[width=0.8\linewidth]{figures/Jang_Figure3.eps}
    \caption{Dimensionless rates $\kappa= \hbar\sqrt{k_BT\lambda}/(\sqrt{\pi}J^2)k_{_{\rm F}}$ (in natural logarithmic scale) versus \rc{-$\Delta G_{jf}/\lambda$ (subscript $jf$ is dropped in the axis label for convenience)} for the case where the spectral density of Eq. (\ref{eq:bath-sp}) is given by an Ohmic form with exponential cutoff,  ${\mathcal J}(\omega)=\pi\hbar \eta \omega e^{-\omega/\omega_c}$ with $\eta=1$ and $\hbar\omega_c=5k_BT$.  $\lambda=5k_BT$ for this case.  Results based on Eqs. (\ref{eq:kf-m}) (Marcus), (\ref{eq:kf-2-sca-1}) \rc{(SCA)}, and (\ref{eq:kf-2-spi}) (SPI) are compared with exact numerical evaluation of ${\rm k_{_F}}$.  }
    \label{fig:spa}
\end{figure}

 While Eq. (\ref{eq:kf-2-sca-1}) works well for negative $\Delta G_{jf}$ with magnitude comparable to $\lambda$,\cite{siders-jacs103-2} it was shown to have significant errors\cite{siders-jacs103,bader-jcp93} for \rc{$\Delta G_{jf} \sim 0$} and may even be worse than the classical Marcus expression, Eq. (\ref{eq:kf-m}) (see Fig. \ref{fig:spa}).   This is because Eq. (\ref{eq:kf-2-sca-1}) is a stationary phase approximation for Eq. (\ref{eq:kf-3}) around a particular value of \rc{$\Delta G_{jf}=-\lambda$}, an approximation that becomes invalid for other values of energy gap.   For the case of  \rc{$\Delta G_{jf} \sim 0$}, the following alternative expression\cite{vanduyne-cp5,siders-jacs103,lang-cp244} was known to be more appropriate.
 \ben
k_{_{{\rm F},j\rightarrow f}}^{\rm SPA-1}&=&|J_{jf}|^2 \sqrt{\frac{\pi\beta}{\hbar^2\lambda_{qs}}}  \exp\Big\{ -\beta \frac{(\Delta G_{jf}+\lambda_{qs})^2}{4\lambda_{qs}} \nonumber \\
&&\hspace{1.2in}+\frac{\beta}{4}(\lambda_{qs}-\lambda_{qt})  \Big \}  , \label{eq:kf-2-spa-1}
\een
where $\lambda_{qs}$ and $\lambda_{qt}$ are quantum reorganization energies employing $\sinh$ and $\tanh$ functions respectively and are respectively defined as 
\ben
&&\lambda_{qs}\equiv \frac{\beta\hbar^2}{2}\sum_n \delta g_{n,jf}^2 \frac{\omega_n^2} {\sinh(\frac{\beta\hbar\omega_n}{2})} \nonumber \\
&&\hspace{.3in}=\frac{\beta\hbar}{2\pi}\int_0^\infty d\omega {\mathcal J}_{jf}(\omega) \frac{1}{\sinh(\frac{\beta\hbar\omega}{2})} ,  \\
&&\lambda_{qt}\equiv \frac{4}{\beta}\sum_n \delta g_{n,jf}^2 \tanh(\frac{\beta\hbar\omega_n}{4}) \nonumber \\
&&\hspace{.3in}=\frac{4}{\pi\beta\hbar}\int_0^\infty d\omega \frac{{\mathcal J}_{jf}(\omega)}{\omega^2} \tanh(\frac{\beta\hbar\omega}{4})  .
\een 
 Later, Jang and Newton\cite{jang-jpcb110} provided a comprehensive analysis of all the stationary phase approximations including the case \rc{$\Delta G_{jf} \sim \lambda$}, for which 
 the analogue of Eq. (\ref{eq:kf-2-sca-1})  becomes
\ben
k_{_{{\rm F},j\rightarrow f}}^{\rm SPA-2}&=&|J_{jf}|^2 \sqrt{\frac{\pi\beta}{\hbar^2\lambda_{qc}}}  \exp\Big \{ -\beta \frac{(\Delta G_{jf}+2\lambda_{qc}-\lambda)^2}{4\lambda_{qc}}  \nonumber \\
&&\hspace{1.2in}+\beta(\lambda_{qc}-\lambda_{q})  \Big \}   . \label{eq:kf-2-spa-2}
\een 
\rc{In the  classical or high temperature limit of  $\beta\hbar\omega_n \ll1$ for all $\omega_n$, both Eqs. (\ref{eq:kf-2-spa-1}) and (\ref{eq:kf-2-spa-2}) also approach the Marcus expression, Eq. (\ref{eq:kf-m}), confirming the universality of the latter expression.}

It is important to note that Eqs. (\ref{eq:kf-2-sca-1}), (\ref{eq:kf-2-spa-1}), and (\ref{eq:kf-2-spa-2}) are valid for only narrow ranges of $\Delta G_{jf}$, which can be improved if more accurate values for the pre-exponential factors are used.\cite{jang-jpcb110}  A unified expression interpolating these three stationary phase expressions has also been derived and is given by\cite{jang-jpcb110} 
\be
k_{_{{\rm F},j\rightarrow f}}^{\rm SPI}=|J_{jf}|^2 \sqrt{\frac{\pi\beta}{\hbar^2\lambda_{qc} U_{jf}}}  \exp\left \{ -\beta \frac{(\Delta G_{jf}+\lambda)^2}{4\lambda_{qc}}  \right\}  , \label{eq:kf-2-spi}
\ee
where
\be
U_{jf}=\left [\left (1- \frac{\lambda_{qs}}{\lambda_{qc}} \right )\left (\frac{\Delta G_{jf}}{\lambda} \right)^2+\frac{\lambda_{qs}}{\lambda_{qc} }\right ] \left (1+\alpha e^{\gamma \Delta G_{jf}/\lambda}\right) ,
\ee
with 
\ben
&&\alpha=\exp\left \{\beta \frac{\lambda_{qc}\lambda_{qt} -\lambda^2}{2\lambda_{qc}}\right\} -1 ,\\
&&\gamma=\ln \left [ \exp\left\{ 2\beta \frac{\lambda (\lambda_{qc}-\lambda)}{\lambda_{qc}} \right\} -1 \right ]-\ln \alpha . 
\een

Comparison of Eq. (\ref{eq:kf-2-spi}) with exact numerical values (see Fig. \ref{fig:spa}) shows its good performance for \rc{$\Delta G_{jf} \geq -2\lambda$}.  Although this expression is not universally applicable, conditions for its validity are well defined\cite{jang-jpcb110} and model calculations\cite{,jang-jpcb119,chen-jcp153} so far confirmed its reasonable performance for the case \rc{$|\Delta G_{jf}|$} is on the order of $\lambda$.  Applications to the hole transport in polyethylene\cite{sato-itdei23-2} also demonstrated the effectiveness of Eq. (\ref{eq:kf-2-spi}).

\subsection{Analytic continuation to imaginary time and instanton calculation}
The equivalence between the equilibrium quantum correlation function and its analytic continuation to the imaginary time was well established\cite{baym-jmp2} and has often been used to bypass the numerical sign problem of the former.  
For the evaluation of FGR, this means evaluating the KT's generating function, Eq. (\ref{eq:gen-ke}).   One attractive method to evaluate this is to use the instanton,\cite{callan-prd,coleman-prd,affleck-prl46,benderskii-pr233,cao-jcp103,cao-jcp106,schwieters-jcp108,jang-jcp114} which amounts to quadratic expansion around the stationary paths in the path integral representation of  the imaginary time correlation function.  

In recent years, Richardson and coworkers developed a series of methods extending the instanton approach for the calculation of FGR.\cite{richardson-jcp143-1,richardson-jcp143-2,richardson-jpcl15,heller-jcp152-1,heller-jcp152-2,ye-jctc21}  These instanton theories include the stationary phase approximation for the harmonic oscillator bath model as specific limits and can furthermore account for anharmonic and curvature effects more generally.   Implementation of these methods is nontrivial because instanton trajectories satisfying proper boundary conditions have to be identified numerically, but  are expected to serve as unique approaches for calculating FGR rates.

 \section{Applications and generalizations of FGR}
FGR  has been applied  to a broad range of quantum processes, which can be classified according to the type of coupling Hamiltonian $\hat H_c$ and the nature of initial and final states. Our focus here will be mostly  on processes involving electronic transitions in molecular systems where quantum mechanical treatments are essential. 
When appropriate, we will address more specialized approximations and assumptions employed for each application. 

\subsection{Radiative transitions and calculation of spectroscopic observables}
\subsubsection{Application of energy domain expression \label{radiative-energy}}
Radiative transitions in steady states constitute the most direct and well known applications of FGR and include standard spectroscopies such as absorption, stimulated emission, Raman scattering, and spontaneous emission. 
The elementary FGR rate expression used for calculating these spectroscopic transition rates  is the following energy domain expression:
\be
k_{_{{\rm F},j\rightarrow f}}(\omega)=\frac{2\pi}{\hbar}\left |\langle \psi_{f}|\hat H_{mr}|\psi_{j}\rangle \right |^2 \delta (E_{m,j}-E_{m,f}\pm \hbar\omega)  , \label{eq:kfgr-sp-e}
\ee
where $E_{m,j}$ and $E_{m,f}$ are energies of initial and final molecular states (including their environments), $\hat H_{mr}$ is the matter-radiation interaction Hamiltonian, and $\omega$ is the angular frequency of the photon being absorbed ($+$) or emitted ($-$). 

Within the semiclassical approximation for the radiation with a stationary frequency, which is sufficient for describing standard absorption and scattering spectroscopies, Eq. (\ref{eq:kfgr-sp-e}) can be obtained from  Eq. (\ref{eq:kfgr-time-s}) with $\hat H_c(t)=\hat H_{mr}e^{- i\omega t}+\hat H_{mr}^\dagger e^{i\omega t}$, for which $|\psi_j\rangle$ and $|\psi_f\rangle$ represent the molecular states (including their environments). In  more rigorous quantum electrodynamics formulation of the radiation,\cite{fermi-rmp4,fermi-rmp4,craig}  which is needed for microscopic derivation of spontaneous emission spectra, Eq. (\ref{eq:kfgr-sp-e}) is obtained by including appropriate quantum radiation states  into $|\psi_j\rangle$ and $|\psi_f\rangle$.   
For traditional far-field spectroscopies, the matter-radiation  interaction $\hat H_{mr}$ can be treated within the electric dipole approximation, but going beyond such approximation is essential for near-field spectroscopies\cite{dunn-cr99,jain-pnas109} or for transitions due to high energy radiation.\cite{list-jcp142,brumboiu-jctc15} 

With advances in computational methods, accurate calculation of the matter-radiation coupling terms and major states involved in transitions have become possible even for complex systems.  This has enabled quantitative prediction/modeling of spectroscopic observables based on FGR.   A prime example  is the molecular absorption spectroscopy, for which vibrationally resolved optical spectra can be calculated in a routine manner explicitly including the initial and final vibronic eigenstates. Within the Condon approximation, major features of the spectroscopy can be calculated only through  determination of Franck-Condon factors, for which various methods have been well established\cite{kubo-ptp13,markham-rmp31,sharp-jcp41,mebel-jpca103,jankoviak-jcp127,Guo2012FranckCondon} within the harmonic approximations for all the vibrational modes. In recent years, significant theoretical advances that account for Duschinsky rotation,\cite{mebel-jpca103,chang-jcp128,ianconescu-jpca108,borrelli-jcp129,santoro-jcp128,peng-cp370,Guo2012FranckCondon,bloino-jcp128}  non-Condon,\cite{huh-jpcs380} Herzberg-Teller,\cite{santoro-jcp128,peng-cp370,Guo2012FranckCondon} and anharmonic effects\cite{harshbarger-jcp53,wang-jpca113,huh-jpcs380,bloino-jcp128} have been made.  Many of these employed the KT generating function approach\cite{kubo-ptp13} that can be also viewed as a time domain approach for finite temperature and equilibrium distribution.   Further extension of these FGR level theories by cumulant approximations\cite{anda-jctc12,zuehlsdorff-jcp151,wiethorn-jcp159} have also been made. 

FGR has also provided key mechanistic understanding of modern X-ray and photoelectron spectroscopies. For example, X-ray absorption spectroscopy  is routinely formulated as a dipole-induced transition rate from a core level to unoccupied final states, while subsequent core-hole decay channels, such as radiative emission or Auger decay, can be described by FGR rates for respective processes.\cite{Xray}  For these, FGR provides clear connections between microscopic transition operators and measured steady-state or quasi-steady-state signals.

For spontaneous emission and fluorescence spectroscopies, where 
excited electronic states decay through photon emission into the background electromagnetic modes,
application of FGR with explicit account of quantum states of radiation provides fundamental understanding of how radiative decay is modulated by the photonic density of states. 
This is particularly important for nonnatural environments such as inside a cavity and in heterogeneous environments. 
FGR can indeed describe cavity-induced enhancement or suppression of radiation, namely the Purcell effect,\cite{Purcelleff} in the limit of weak light-matter coupling.  Even for cases where there are strongly coupled plasmons or photonic states, forming polaritons,  FGR can be used with proper definitions of hybrid states as initial and final states.     

\subsubsection{Application of time domain expression for spectral lineshape and time resolved spectroscopy}
Accurate calculation of spectral lineshapes in general entails challenging or unresolved theoretical  issues\cite{breene} that often go beyond the assumptions of FGR.  
However, even in the limits where  the assumptions {\bf C1} and {\bf C2} validating FGR are met, accurate modeling of experimental lineshapes is challenging. Employing the energy domain expression for FGR typically involves introducing phenomenological lineshape functions dressing the Dirac-delta function part of Eq. (\ref{eq:kfgr-sp-e}).  However, this practice lacks clear physical justification except for very simple situations.  

For molecular systems, there are multitudes of line-broadening mechanisms ranging from fast electronic decoherence to  slow conformational fluctuations and environmental responses.  These are in general interconnected with each other and difficult to disentangle.    To account for these effects,  the time domain expression, Eq. (\ref{eq:kfgr-time-s}), serves as a more natural framework especially for complex systems in condensed environments,\cite{mukamel,Cho2009book}  since the notion of well separated peaks with independent broadening mechanisms, as implicit in Eq. (\ref{eq:kfgr-sp-e}), is no longer valid.  In fact, lineshape calculation is an important test for the accuracy of quantum dynamics theories and Hamiltonians representing molecular systems and their interactions with environments.  

 While exact quantum dynamics calculations remain daunting,  there are now well-established  approximate methods such as semiclassical propagation of wave packets\cite{heller-jcp68,heller-acr14,heller-acr39,wehrle-jcp140,begusic-jcp150,begusic-jcp153,zhang-jctc21,cina} or Wigner distributions.\cite{yan-jcp88,mukamel,mcrobbie-jpca113,kwac-jpcb112}    For systems in condensed and/or complex environments, it is often necessary to use open system quantum dynamics approaches to account for the effects of infinitely many degrees of freedom.    Various approaches\cite{mukamel,cina,jang-exciton,tanimura-jpsj75,banchi-jcp138,schroder-jcp124,zhang-jcp108,yang-cp275,yang-jcp123,gelzinis-jcp142,dinh-jcp142,dinh-jcp145,song-jcp143,jang-jcp151,schulz-cr124} have been tested and developed successfully to this end.

Time domain expression for FGR can also be easily generalized for nonequilibrium situations  such as pump-probe spectroscopy.\cite{mukamel,cina} 
In contrast to steady-state measurements, spectroscopic observables for these cases depend explicitly on finite duration of  probing that follows excitation.  In particular for an impulsive pulse, its effects can be represented by a nonequilibrium initial state either in wavepacket\cite{heller-acr14,heller-acr39} or Liouville space\cite{mukamel,pumpprobe} formulations.   Transitions due to a delayed probe pulse, which can be described at the level of Eq. (\ref{eq:kif-time-1}), then can be used with finite $t_s$  and appropriate time-dependent coupling Hamiltonian $\hat H_c(t)$. 
This formulation naturally captures oscillatory features arising from coherent dynamics, while accounting for broadening of signals due to finite duration of probe pulses, decoherence, ensemble dephasing and environmental interactions. These nonequilibrium FGR calculation approaches can also be combined with high-level {\it ab initio} calculation methods for quantitative modeling of emission spectra.\cite{PAH07}

\subsection{Charge transfer processes}
Charge transfer (CT) processes have played central roles in earlier applications of FGR.  In particular, research on ET has led to advances in computational methods for applying FGR to molecular systems in liquids and complex media.    A comprehensive review of charge transfer processes is beyond the scope of this work, for which there is already authoritative  literature.\cite{marcus-arpc15,marcus-bba811,Ulstrup,newton-cr91,adv-et,newton-tca110,blumberger-cr115,hammes-schiffer-cr110,migliore-cr114} We here provide brief accounts of some select topics of CT processes directly relevant to the present perspective of FGR. 

\subsubsection{Quantum nonadiabatic ET theories}
Major advances in theories\cite{marcus-jcp24,marcus-jcp43,levich-dan124} of ET began in late 1950s with new experimental advances in radiation chemistry and  electrochemistry.  While Marcus laid the most well recognized foundation of ET through a series of celebrated papers,\cite{marcus-jcp24,marcus-jcp26-1,marcus-jcp26-2,marcus-dfs29,marcus-jpc67,marcus-jcp43,marcus-arpc15,marcus-bba811,marcus-jcp81} his earlier theories\cite{marcus-jcp24,marcus-jcp26-1,marcus-jcp26-2,marcus-dfs29,marcus-jpc67,marcus-jcp43} were \rc{mostly based on the assumption of activated processes over adiabatic potential energy surfaces while accounting for some nonadiabatic effects.  Hush\cite{hush-jcp28,hush-tfs} also developed similar adiabatic theories.  On the other hand,} Levich and Dogonadze\cite{levich-dan124,ulstrup-rje53} were the first who developed the formalism of  nonadiabatic ET by applying FGR.  
This theory is closely related to the theory of nonradiative decay by KT\cite{kubo-pr86,kubo-ptp13} and Holsten's polaron theory\cite{holstein-ap8-2} in the weak coupling limit.   \rc{Later}, Kestner {\it et al.}\cite{kestner-jpc78} and Ulstrup and Jortner\cite{ulstrup-jcp63}  developed more comprehensive quantum formulations where the ET process can be viewed as a nonradiative decay of a super-molecular system and made it natural to define the corresponding FGR as the nonadiabatic ET rate.    Alternative quantum formalism within the single electron approximation have also been made.\cite{onuchic-jpc90}

If all the modes in the quantum formulation of ET can be modeled by displaced harmonic oscillators with continuous bath spectral densities, quantum formulations\cite{kestner-jpc78,ulstrup-jcp63,onuchic-jpc90} reduce to the spin-boson model\cite{leggett-rmp59,garg-jcp83} for which the semiclassical or stationary phase approximations described in Sec. IV.A can serve as good approximations for appropriate values of reaction free energies.   On the other hand, for cases where few vibrational modes make major contributions, Jortner\cite{jortner-jcp64} derived more refined quantum rate expression for ET that has played an important role for the modeling of ET involving biological and organic molecules.    

Further generalizations of spin-boson type model to account for non-Condon effects,\cite{daizadeh-pnas94,medvedev-jcp107,medvedev-cp296,troisi-jcp119,berlin-jpcc112,jang-jcp122,sun-jcp144} anharmonic contributions,\cite{islampour-cpl179,yeganeh-jcp124,schmidt-jcp138} and nonlinear couplings\cite{tang-cp188,matyushov-jcp113,freed-jpcb107} were developed.  For multistate ET processes, calculation of electronic coupling constants through multiple pathways\cite{newton-tca110,adv-et,newton-tca110,beratan-jcp86,beratan-science258,balabin-science290,stuchebrukhov-jcp104,stuchebrukhov-jcp105} has been a primary issue, for which various computational advances have been made.  Many of these important effects were  described well by FGR level theories in combination with advanced quantum calculations and statistical sampling methods.  Even for photo-induced ET processes, nonequilbrium extensions\cite{coalson-jcp101} of FGR, when  combined with advanced quantum dynamics methods,\cite{sun-jctc12,kananeka-jcp148,sun-jcp145,hu-jpcb124} were shown to be successful for describing a broad range of seemingly complex data with near quantitative accuracy.     

\subsubsection{Proton and proton coupled electron transfer theories}
Proton transfer (PT) reactions have been widely studied in the context of acid-base reaction,\cite{weiss-jcp41} hydrogen bonding, kinetic isotope effects,\cite{heeb-jpcb112} and catalysis.  A broad class of slow PT processes can be treated within the framework of the quantum transition state theory that accounts for tunneling,\cite{weiss-jcp41} where the reaction coordinate involves proton movement. On the other hand, the similarity of environmental response to PT with that for ET has made it possible to extend various theories of ET  for PT processes. For example, Marcus extended his theory to PT in the context of acid base reactions.\cite{marcus-jpc72}  
Levich, Dogonadze, Kuznetzov, Ulstrup, and  their coworkers developed FGR-level theories\cite{levich-ea15,vorotyntsev-dan209,german-jcsft76,german-jcsft77,kusnetzov-ea32} for PT in the nonadiabatic regime, where adiabatic separation of proton state from other nuclear degrees of freedom can be made.     

For molecular systems, Cukier and coworkers developed more comprehensive FGR-level theories\cite{cukier-jcp91,morillo-jcp92} of PT based on detailed model Hamiltonians while also accounting for contributions of polarization responses systematically.\cite{cukier-jpcb101} Hynes and coworkers\cite{borgis-cpl89,borgis-cp170,borgis-jcp94,borgis-jpc100,kiefer-jpca106-1,kiefer-jpca106-2,kiefer-jpca108-1,kiefer-jpca108-2,kiefer-jpcb125} also developed a comprehensive series of PT theories encompassing both nonadiabatic and adiabatic regimes, and established computational methods to extract time correlations involved in their FGR-type rate expressions directly from MD simulations.  Applications of these theories clarified key factors contributing to PT reactions under different situations and the criteria that make application of FGR-type theory  appropriate. 

Many PT processes are coupled to ET reactions, especially in biological systems, where protein residues provide ample resources for loose but localized electrons and protons. Since the quantum nature of protons become amplified for these proton-coupled electron transfer (PCET) reactions, significant theoretical developments were needed for both mechanistic understanding and quantitative modeling of rates.  Cukier\cite{cukier-jpc98,cukier-jpc99,cukier-jpc100,cukier-jpca103,cukier-jpcb106}  identified important characteristics of PCET and established theoretical frameworks for describing different mechanisms of PCET based on FGR.  These include the importance of non-Condon effects\cite{cukier-jpc98,zhao-jpc99} when protons participate directly in modulating the coupling for electron transfer, and classification of PCET depending on  whether proton and electron move in concerted or sequential manner.  For concerted PCET where there is significant protein tunneling,  Georgievskii and Stuchebrukhov\cite{georgievskii-jcp113} also developed a FGR-level theory (for the transferring electron) while accounting for proton tunneling explicitly along the reaction coordinate.  

Despite theoretical advances, quantitative modeling and predictions of PCET processes remained challenging, especially in biological systems.  Hammes-Schiffer and coworkers made major advances\cite{hammes-schiffer-cr110,layfield-cr114} addressing these issues through a combination of general theoretical formulation and detailed computational studies.  In particular, Soudackov and Hammes-Schiffer\cite{soudackov-jcp111,soudackov-jcp113} developed a general theories that defines PCET on \rc{two-dimensional} solvent reorganization coordinates and established a unified FGR rate theories that can be applied to a broad range of PCET processes.

\subsection{Spin-preserving nonradiative electronic transitions}

Spin-preserving nonradiative electronic transitions have long been described by FGR. These  include intramolecular internal conversion (IC), EET, and the initial electronic step of singlet fission (SF). All of these involve transitions between molecular quantum states of the same spin multiplicity coupled by perturbation Hamiltonians that do not explicitly involve spins. In fact, if we adopt a more general super-molecule perspective, all of these may be viewed as different realizations of IC for the group of molecules involved in the transition.  However, we here employ the conventional view that IC refers to intramolecular transitions.  Applications of FGR for these processes were indispensable for analyzing nonradiative relaxation processes across diverse systems such as dye molecules in complex environments,  molecular liquids, and wide-bandgap semiconductors.  In conjunction with experimental efforts, such applications also enabled  optimizing material properties toward advancing the development of high-performance devices.  More detailed \rc{accounts} of the three major processes are provided below.

\subsubsection{Internal conversion (IC) between electronic states}
The primary mechanism for IC between electronic states is through nonadiabatic couplings (NACs) between adiabatic electronic states, which exist due to the breakdown of Born-Oppenheimer approximation and involve derivatives of adiabatic electronic states with respect to nuclear coordinates.  Although this is a well known feature in molecular quantum mechanics, subtle or complicated issues concerning its calculation and approximations in the context of rate calculation have motivated various theoretical formulations and studies over many decades.\cite{kubo-pr86,kubo-ptp13,lin-jcp44,lin-jcp58,robinson-jcp37,bixon-jcp48,sharf-cpl9,orlandi-cpl8,nitzan-jcp56,fujimura-jcp66,mebel-jpca103,niu-jpca114,ICrate,IC,jang-jctc21}   Accurate account of NACs has also been a central theme in the development of nonadiabatic quantum dynamics methods.\cite{deumens-rmp66,zhu-jcp137,jasper-acr39,kapral-arpc57,abedi-jcp137,jang-jcp-nonad,tully-jcp137,wang-arpc66,subotnik-arpc67,goings-wcms2017,makhov-cp493,curchod-cr118,joubert-doriol-jpca122,zhao-jpcl11,song-jctc16,esch-jcp155,prezhdo-acr54,guo-jcp155,shu-jctc18,he-wires12,huang-jcp159,he-jpcl15,ICrate}  Despite significant advances in these time dependent dynamics methods, applications of FGR to many IC processes have been well justified and successful, as exemplified by various theories and computational methods.\cite{kubo-pr86,kubo-ptp13,lin-jcp44,lin-jcp58,bixon-jcp48,fischer-jcp53,nitzan-jcp56,fujimura-jcp66,mebel-jpca103,niu-jpca114,peng-cp370,wang-jcp154,borrelli-jcp129,sharf-cpl9,orlandi-cpl8,ICrate,IC,jang-jctc21}  

Early on,  there were numerous applications of FGR to IC processes between electronic states employing the energy domain expression, Eq. (\ref{eq:kfgr-0}) or (\ref{eq:kfgr-g}).  Within these approaches, the IC rates are governed by two key factors: the magnitude of the vibronic coupling matrix elements and the density of final vibronic states that meet the energy conservation requirement. However, resolving the issues of {\bf F1}-{\bf F3} described in Sec. \ref{section2} remained challenging.   For example, issues concerning different ways to employ the final density of states\cite{robinson-jcp37,bixon-jcp48} and to define proper zeroth order and coupling Hamiltonians\cite{sharf-cpl9,orlandi-cpl8} have not been resolved.  

Theories\cite{mebel-jpca103,niu-jpca114,peng-cp370,wang-jcp154,borrelli-jcp129} extending the GF approach\cite{kubo-pr86,kubo-ptp13} or using the integrand of the time domain expression helped resolving the issue {\bf F2}  by allowing effective enumeration of all coupling matrix elements within the harmonic approximation for the vibrations, for which general expressions\cite{mebel-jpca103,niu-jpca114,borrelli-jcp129} including the effects of Duschinsky rotation and Herzberg-Teller couplings have been derived.  However, the issue {\bf F3}, which is critical for evaluating  small rates with reliable accuracy, remains unresolved in these approaches since application of the closed-form expressions still require using line-broadening or apodization function.  Although widely used Lorentzian broadening function may be justified according to Bixon and Jortner,\cite{bixon-jcp48} the underlying assumption of infinite equidistant final vibrational levels assumed in this work is not general enough.    

The main issue underlying {\bf F3} when using Hamiltonian with finite degrees of freedom is the assumption of closed-system unitary dynamics for molecular Hamiltonian.  This issue can be resolved by including a bath Hamiltonian that properly represents all the environmental response and using the time  domain expression Eq. (\ref{eq:kfgr-b-fi}).   The choice of system part in this expression can represent any vibronic states and the bath can also be general.  However, most theories developed so far have assumed the minimal electronic states as system and linearly coupled harmonic oscillator bath, which can be justified within the linear response theory\cite{georgievskii-jcp110} and allows derivation of a closed form expression for the bath contribution. In particular, given that all the molecular vibrational modes and the bath can be approximated as harmonic oscillators with bilinear couplings at most, the normal mode transformation result in a bath of independent harmonic oscillators with continuous spectral density.  Within this model, rate can be defined unambiguously for a broad range of bath spectral densities although care should be taken in some cases.\cite{jang-jcp159,jang-jcp161}     

Experimentally, many IC rates of excited electronic states decrease almost exponentially when the energy gap (EG) of the excited state from the ground/final electronic state becomes large.  Application of FGR in this regime has been particularly important since these are typically  in the time regime inapproachable by real time quantum dynamics methods.  Given that the dominant factor contributing to these slow processes is the Franck-Condon weighted density of states, which is expected for large enough EG, the experimental observation can be modeled by approximating the effects of NACs with \rc{effective constant coupling terms}.  This was the approach taken by Englman and Jortner (EJ) in their development of a seminal EG law (EJ-EG) theory,\cite{englman-mp18,radiationlesstransition} and other related works.\cite{fischer-jcp53,freed-acr11,gelbart-jcp52,freed-jcp52,fong-jcp56,nitzan-jcp63,freed-acr11,lin-jcp58,kober-jpc90}  

The EJ-EG law\cite{englman-mp18} is in fact a particular application of Eq. (\ref{eq:kf-3}) within the stationary phase approximation as described in Appendix \ref{spa}.   For the case where the dominating factor on the right side of Eq. (\ref{eq:sta-eqn}) comes from the highest frequency vibrational modes with its frequency $\omega_h >> k_BT/\hbar$, Eq. (\ref{eq:sta-eqn}) can be approximated as 
\be
\frac{\Delta G_{jf}}{\hbar}\approx -  N_h\delta g_{h,jf}^2 \omega_h \frac{\exp \left (\frac{\beta\hbar\omega_h}{2}-i\omega_h z_s\right) }{\exp \left (\frac{\beta\hbar\omega_h}{2}\right )} ,  \label{eq:sta-eqn}
\ee    
for which the following imaginary value is a solution
\be
z_s=-\frac{i}{\omega_h}\ln \left (\frac{-\Delta G_{jf}}{\lambda_h}\right ) , 
\ee
with $\lambda_h=N_h \delta g_{h,jf}^2 \omega_h$.  

\begin{figure}
    \centering
    \includegraphics[width=0.9\linewidth]{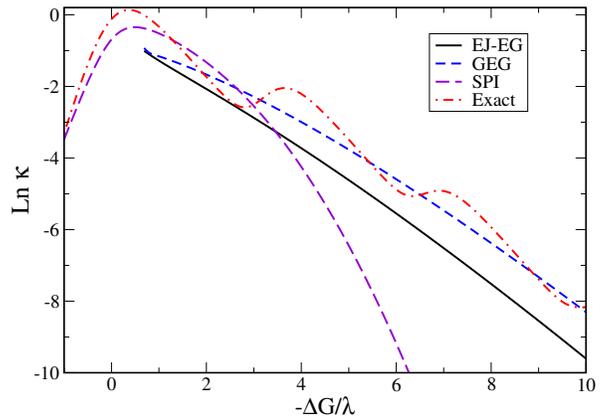}
    \caption{Dimensionless rates $\kappa= \hbar\sqrt{k_BT\lambda}/(\sqrt{\pi}J^2)k_{_{\rm F}}$ (in natural logarithmic scale) versus -$\Delta G/\lambda$ for the case where the spectral density of Eq. (\ref{eq:bath-sp}) is given by ${\mathcal J}(\omega)=\pi\hbar \eta \omega e^{-\omega/\omega_c}+\pi \hbar s_h \omega_h^2 \delta (\omega-\omega_h)$ with $\eta=1$, $\hbar\omega_c=5k_BT$, $s=0.1$, and $\omega_h=5\omega_c$.  Rates calculated by EJ-EG law,\cite{englman-mp18} GEG law,\cite{jang-jcp155-1} and Eq. (\ref{eq:kf-2-spi}) (SPI) are compared with exact numerical evaluation of ${\rm k_{_F}}$. }
    \label{fig:geg}
\end{figure}

 The rate expression of EJ-EG law\cite{englman-mp18} is obtained\cite{englman-mp18,jang-jcp155-1} by using the above value for $z_s$ and corresponding expressions for $f(z_s)$ and $f''(z_s)$ in Eq. (\ref{eq:kf-2-st}) and defining $-\Delta G_{jf}$ as the EG.   This EJ-EG law\cite{englman-mp18} has been widely used and helped quantifying and clarifying various nonradiative processes.  As yet, it has served more as a phenomenological guide rather than a quantitative theory.  For example, a recent work\cite{friedman-chem7} by Caram and coworkers showed that a class of near infrared (NIR) and short wavelength infrared (SWIR) dye molecules indeed follow the trend of EJ-EGL theory\cite{englman-mp18} very well, except that  the effective coupling estimated from the fitting of experimental data was unusually high. This can be explained based on a recent theoretical generalization of the EG law (GEG),\cite{jang-jcp155-1} which  included the contribution of low frequency vibrational modes as well.  the enhancement due to the low frequency vibrational modes can enhance the IC rate systematically while not affecting the EG behavior significantly (see Fig. \ref{fig:geg}).  
 
 Indeed, detailed computational modeling\cite{ramos-jpcl15} of  two NIR and SWIR dye molecules demonstrated that application of FGR including all low frequency vibrational modes can account for quantitative trends of rates including isotope effects.  Recently, more satisfactory theory\cite{jang-jctc21} including non-Condon effects due to nuclear momenta coupled to NAC terms was developed within the harmonic oscillator bath model and quasiadiabatic approximation.  Applications\cite{jang-jctc21,park-jpcl17} of this theory to the nonradiative decay from the first and second excited states of azulene helped clarify dynamical details of its anti-Kasha behavior.        

Even for moderate values of EG where near exponential dependence is not observed, there are many examples for which FGR is expected to work.  
Furthermore, if non-equilibrium versions of FGR are used in conjunction with MEs as described in Sec. \ref{me-fgr}, quantitatively accurate description of IC processes is possible even when the time scale for IC processes is comparable to time scales of vibrational relaxation.   Of course, as the energy gap decreases further or NACs becomes too large as in near conical intersections (CIs), the underlying assumptions justifying FGR no longer hold.  Indeed, strong perturbations associated with CIs often lead to very fast IC.\cite{UfastCI-18} However, even in such cases, nonequilibrium FGR theory developed by employing a proper electronic basis\cite{izmaylov-jcp135} can offer reasonable description of  ultrafast IC dynamics with reasonable accuracy. 

Although theoretical and computational advances that have been made for several decades now enable calculation of FGR rates for IC rates for a broad range of systems, there are long standing theoretical issues that require more advanced theories and computational methods.   These include non-Condon effects\cite{nitzan-jcp56} arising due to the dependence of adiabatic electronic states on nuclear coordinates, the effects of anharmonic vibration, for which various new computational approaches are being developed,\cite{Makri-anhar-2024}  and the full effects of the second derivative NAC terms that are often ignored but can have nontrivial effects.

\subsubsection{Excitation energy transfer}    
EET\cite{forster-ap,forster-book,dexter-jcp,agranovich,eet-rev,silbey-arpc27,ret,may,olaya-castro-irpc30,jang-exciton} is the transfer of excitation through quantum mechanical resonance interactions that do not require physical contact, and is thus also called resonance energy transfer (RET).\cite{forster-book,dexter-jcp,ret,jang-exciton}  This is mostly nonradiative but can occur radiatively at long distances, for which explicit consideration of quantized radiation field is needed.\cite{juzeliunas-prb49,juzeliunas-prb50} Most theoretical and experimental studies of EET have been on transfer of electronic states, but resonant transfer of vibrational states is also possible and has been studied.\cite{yang-jcp135}  Our main focus here is the nonradiative transfer of electronic states that constitute the majority of EET processes.      
 
 The initial and final states underlying EET involve direct product states $|j\rangle=|j_{_D}\rangle \otimes |j_{_A}\rangle$,  $|k\rangle=|k_{_D}\rangle \otimes |k_{_A}\rangle$. Here, $|j_{_D}\rangle$ and $|k_{_D}\rangle$ are electronic states of the energy donor (D), with the former having higher energy, whereas $|j_{_A}\rangle$ and $|k_{_A}\rangle$ are electronic states of the energy acceptor (A), with the latter having higher energy.   Thus, the excitation for D corresponds to $|k_{_D}\rangle \rightarrow |j_{_D}\rangle$ transition, whereas that for A is $|j_{_A}\rangle \rightarrow |k_{_A}\rangle$.  The standard assumption is that $|j\rangle$ and $|k\rangle$ are orthogonal to each other, which can be justified as long as the distance between D and A is large enough not to form chemical bond.  
Two well-known mechanisms of EET are Coulomb\cite{forster-ap,forster-book,dexter-jcp} and exchange\cite{dexter-jcp} interactions.  

The Coulomb interaction term between the states $|j\rangle$ and $|k\rangle$ is defined as $\langle j|\hat H_{_{\rm Coul}}|k\rangle$, where $\hat H_{_{\rm Coul}}$ is the  Coulomb potential operator between electrons in D and A.  This term does not vanish as long as  $|k_{_D}\rangle \rightarrow |j_{_D}\rangle$ and $|j_{_A}\rangle \rightarrow |k_{_A}\rangle$ are both spin allowed, for which it can also be expressed in terms of products of transition densities involving orbitals only.  Thus, the EET rate due to Coulomb interaction can be calculated by applying FGR for  $|j\rangle$ and $|f\rangle$ as initial and final states and $\hat H_{_{\rm Coul}}$ as a coupling Hamiltonian.  The leading term of these is the transition dipole, for which F\"{o}rster derived\cite{forster-ap,forster-book} a seminal expression known as F\"{o}rster's RET (FRET) rate that can be expressed  as\cite{jang-jcp127,jang-exciton} 
\be
k_{_{\rm FRET}}=\frac{9000 (\ln 10)\langle \kappa_{jk}^2 \rangle}{128 \pi^5 N_A \tau_D n_r^4 R_{DA}^6}\left ( \int d\tilde \nu \frac{f_D(\tilde \nu) \epsilon_A (\tilde \nu)}{\tilde \nu^4}\right) \comma \label{eq:kforster}
\ee
where $\tilde \nu$ is the wavenumber (in the unit of ${\rm cm^{-1}}$), $f_D(\tilde \nu)$ is the normalized emission spectrum of D for the $|j_{_D}\rangle \rightarrow |k_{_D}\rangle$ transition, $\epsilon_{_A} (\tilde \nu)$ is the molar extinction coefficient of A for the $|j_{_A}\rangle \rightarrow |k_{_A}\rangle$ transition, $R_{DA}$ is the distance between the donor and the acceptor,  $n_r$ is the refractive index of the medium, $N_A$ is the Avogadro's number, $\tau_D$ is the spontaneous decay lifetime of D (for the $|j_{_D}\rangle \rightarrow |k_{_D}\rangle$ transition), and $\langle \kappa_{jk}^2 \rangle$ is the average orientational factor of the transition dipole interaction, which is often taken to be $2/3$.

In deriving Eq. (\ref{eq:kforster}), F\"{o}rster combined\cite{forster-ap,forster-book} the standard FGR (derived in a careful manner) with Einstein's relations for absorption and spontaneous emission.  This is also explained in detail in Dexter's comprehensive theoretical work,\cite{dexter-jcp} where EET rate expressions due to transition dipole and quadrupole interaction are also provided. Equation (\ref{eq:kforster}) has made great impacts over many decades in photo-chemistry and photo-physics,\cite{birks,ret-pr,scholes-arpc54} optoelectronics,\cite{ret-pr,may,scholes-arpc54} and structural determination\cite{stryer-pnas58,haugland-pnas63,rasnik-acr38,sahoo-jppc12,dimura-cosb40,jares-erijman-nb21,hellenkamp-nm15} of biological molecules since it allowed determination of rate or distance employing standard spectroscopic data.  The integral of emission and molar extinction in Eq. (\ref{eq:kforster}) includes all the effects of  the Franck-Condon overlap, energy conservation condition, and the final density of states.  The only assumption underlying this is that the molecular vibrations and environmental responses coupled to D and A are independent.   While this is a reasonable approximation for EET processes between well-separated D and A in solution or disordered media, its error can be significant if there are common or shared vibrational modes coupled to two excitations as in crystalline environments or for cases where donor and acceptor form parts of the same large molecule.  Theories accounting for these effects have been developed. A recent computational study also clarified that the commonly shared vibrational modes in molecular dyads can explain why the FRET rate directly estimated from single molecule measurements significantly deviated from Eq. (\ref{eq:kforster}).
  
If the transitions $|k_{_D}\rangle \rightarrow |j_{_D}\rangle$ are $|j_{_A}\rangle \rightarrow |k_{_A}\rangle$ are spin forbidden, the Coulomb interaction term as described above Eq. (\ref{eq:kforster}) vanishes.  However, Dexter\cite{dexter-jcp} showed that transfer of such excitations is still possible because matrix elements  of the Coulomb operator with respect to anti-symmetrized initial and final states do not vanish due to exchange interaction term.  Dexter also recognized\cite{dexter-jcp} that such exchange interaction can be significant at small but large enough distances ($\sim 1 {\rm nm}$) for which direct orbital overlap can still be assumed to be zero.  This exchange mechanism has been able to account for the  transfer of many triplet excited states (for singlet ground states).  For this reason, Dexter transfer became almost synonymous with triplet transfer although actual mechanisms can also be due to other processes such as through charge transfer states.\cite{jortner-jcp42,skourtis-pnas113,bai-jpcc124}   

Since theories by F\"{o}rster\cite{forster-ap,forster-book} and Dexter,\cite{dexter-jcp} there have been steady advances in theories and computational approaches for EET processes.\cite{agranovich,eet-rev,agranovich-hoch,silbey-arpc27}  Key theoretical concepts were developed for calculating electronic coupling constants due to higher order multipolar contributions\cite{silbey-jcp42} and charge transfer states\cite{choi-jcp41} in  broader contexts of molecular excitons.\cite{davydov}  There were extensive theories\cite{gochanour-jcp70,loring-jcp76,huber-prb20,blumen-jcp72,jang-jcp102} and computational modeling\cite{bardeen-arpc65,ahn-cpl446} on exciton transport where F\"{o}rster-Dexter theories were used as transition probabilities to understand and model luminescence properties of solid and liquid materials. There were also earlier \rc{considerations} of nonequilibrium effects in the energy domain, which were called hot transfer.\cite{tekhver-jetp42,sumi-prl50}   

During recent three decades, new kinds of theoretical and computational advances were made\cite{jang-wires3,scholes-arpc54} that offer more quantitative description of EET processes, along with advances in spectroscopies and computational methods.  For example, FRET was generalized to multichromophoric systems,\cite{sumi-jpcb103,kakitani-jpcb103,scholes-jpcb104,jang-prl92,jang-jpcb111} where D and A consist of aggregates of strongly coupled molecules.  Nonequilibrium generalization of FRET,\cite{jang-cp275} where the emission profile in the spectral overlap integral is replaced with stimulated emission profile, was developed, and theory of inelastic FRET was developed according to which some of transferring energy can exchange with quantum degrees of freedom modulating the electronic coupling, resulting in a rate expression that involves a two-dimensional spectral overlap integral.  There were also significant advances in computational methods\cite{krueger-jpcb102,scholes-arpc54,renger-pr102,jang-rmp90,hestand-cr118,ET-rev} to calculate electronic couplings and the effects of environments.\cite{hsu-jpcl8,ding-jcp146,ding-jpcc122,wu-jpcl9,du-cs9}   Advances in quantum dynamics method also enabled more quantitative analyses of FGR level theories of EET.

\subsubsection{Singlet fission}

Singlet fission\cite{smith-cr110,smith-arpc64} is a spin-preserving process where one photoexcited singlet exciton eventually generates two separate triplet excitons. Thus, the coupling Hamiltonian $\hat H_c$ involves spin-orbit interactions.  However, note that the final state still has the same spin symmetry as the initial state. The process proceeds via an intermediate correlated triplet-pair state ($^1$(TT)), and FGR can be used to calculate the microscopic rate for the transition from localized S$_1$ to $^1$(TT). The initial and the final states in this case are direct product states of electronic states of individual molecules, which constitute the diabatic basis of the total system, and the naturally ensuing off-diagonal elements formed by the total electronic Hamiltonian in those diabatic states constitute $\hat H_c$.\cite{Herbert-exciton} One added complication is the fact that charge transfer (CT) states may get involved. Indeed, it has widely been discussed that CT-mediated superexchange effect is important for singlet fission.\rc{\cite{smith-cr110,DHKim-CT-jacs}}

FGR can capture the microscopic pictures of transitions between the initial singlet state and the correlated triplet-pair state, including both direct conversion and indirect superexchange pathways. For reliable predictions of actual systems, properly modeling the phonon coupling aspect in relation to the density expression of the FGR rate is needed. With FGR, experimental trends have been rationalized such as faster fission rates in systems with smaller singlet-triplet energy gaps and stronger intermolecular coupling by employing series of acenes and their derivatives, sometimes with quantitative agreements in the observed rates.\rc{\cite{smith-cr110,smith-arpc64}}

\subsection{Spin-crossing nonradiative transitions}

\subsubsection{Spin crossover in metal complexes}

Spin crossover (SCO) involves transitions between high-spin (HS) and low-spin (LS) configurations in response to external stimuli such as temperature, pressure, or light. These transitions are highly relevant for magnetic sensors, molecular switches, and information storage.
When the HS--LS energy gap is sufficiently small, relatively minor external perturbations can drastically change the relative populations of the states,\cite{SCO00,SCO94} although optical perturbation are required for systems with larger energy gaps. The detailed understanding of the relevant spin relaxation processes is critical for utilizing the SCO properties of materials for 
practical applications. As SCO involves transitions between states of different spin multiplicities, they are typically mediated by strong spin-orbit coupling.  Characterizing spin crossover systems has often adopted FGR to predict the lifetimes of spin states and to understand the paths along which the spin state switching takes place.\cite{Hauser91a,Hauser91b,Jortner80,Marian13} For example, in transition metal complexes, FGR can quantify the rate at which the population transfers from one spin manifold to another, following photo-excitation or thermal activation. Of course, care must be taken in regimes where perturbative assumptions break down.\cite{metal}

Depending on the coupling strength and the effective barrier height dictated by the crossing point, the timescale of SCO dynamics can span from femtoseconds to nanoseconds or even longer. 
For processes involving transition metal complexes, the density of vibrational states near resonance will also be one of the primary factors for determining whether the crossover is ultrafast as in the widely studied iron(II)--tris-(bipyridine), $\rm{ [Fe^{II} (bpy)_3 ]^{2+} }$,\cite{Febpy14,Chergui09,Chergui10,Marian13,Chergui07} 
 or relatively slow.\cite{fgrRates22} FGR thus provides important insights into the connection between the characteristics of electronic states at the metal center and the observed spin dynamics. Recent ultrafast spectroscopic studies document such spin-state dynamics fairly well.\cite{Chergui15rev}

\subsubsection{Intersystem crossing in organic molecules}

The time-domain formulation of FGR is also important for describing intersystem crossing
(ISC), where transitions between states of different spin multiplicities are induced by
weak spin-orbit coupling in conjunction with nuclear motion. In many organic molecular
systems, spin-orbit coupling remains sufficiently weak that ISC and reverse ISC (RISC)
can be treated perturbatively. Under these conditions, time-domain FGR expresses the rates in terms of correlation functions of spin-orbit and spin-vibronic coupling
operators,\cite{niu-jpca114,TCF-2,TCF-3,etinski-jcp134,TCF-5} allowing nuclear and environmental effects to be incorporated explicitly.\cite{kim-jctc16}

According to El-Sayed's rule,\cite{Elsayed,marian-arpc72} ISC is most efficient for transitions involving changes in
orbital characters. When direct spin-orbit coupling between an initial and a final state
is weak, second-order mechanisms involving vibronic coupling within the same-spin
manifold can enhance the transition probability. Such spin-vibronic effects\cite{Marian-spin-vibr} can be
systematically incorporated into the framework of FGR through perturbative
expansions of the electronic states.\cite{expansion}
Additional coupling mechanisms such as nuclear hyperfine coupling may also emerge important.\cite{Friend-spin-vibr,Marian-spin-vibr}  

Sometimes, coupling with vibrational modes plays an important role particularly when there are modes that evolve on timescales comparable to electronic dephasing time, as they may
introduce slowly decaying and oscillatory components into the relevant correlation
functions, thereby affecting finite-time transition probabilities and the inferred rates with numerical significance.\cite{Rossky06,Bittner17} 
While introducing a damping factor for integrating the time domain expression is a common practice,\cite{etinski-jcp134,kim-jctc16} care must be taken with the oscillatory behaviors in the time correlations.

\subsubsection{Applications for characterization of OLED materials}

Once electronically excited, organic light-emitting diode (OLED) materials\cite{burroughes-nature347,kohler-am14,sasaki-nc12,zou-mcf4} relax through diverse photophysical processes and the radiative and nonradiative exciton dynamics directly determine the device efficiencies.
FGR has played a central role in
theoretically analyzing such processes. Under electrical excitation, excitons are generated in singlet and triplet spin manifolds with a 1:3 ratio.
Thus, in purely organic fluorescent emitters, triplet excitons constitute serious loss channels unless they are somehow converted back to singlet excitons, and ISC and RISC govern
exciton harvesting efficiencies. This is the natural reason FGR-based rate expressions have been widely used to evaluate the rates of the involved processes mediated by spin-orbit and spin-vibronic
interactions.\cite{OLED}

For example, thermally activated delayed fluorescence (TADF) materials that utilize RISC from triplet to
singlet states have been studied with FGR formalisms. With the energy gap law, an efficient TADF material typically requires a small
singlet-triplet energy gap ($\Delta E_{\rm{ST}}$), but it should be matched with sufficiently strong
spin-orbit or spin-vibronic coupling. Molecular design strategies based on donor-acceptor
architectures aim to reduce $\Delta E_{\rm{ST}}$ by spatially separating frontier orbitals,
and FGR-based analyses provide a quantitative framework for assessing how such design
choices influence ISC and RISC rates.

\begin{figure}
    \centering
    \includegraphics[width=1.0\linewidth]{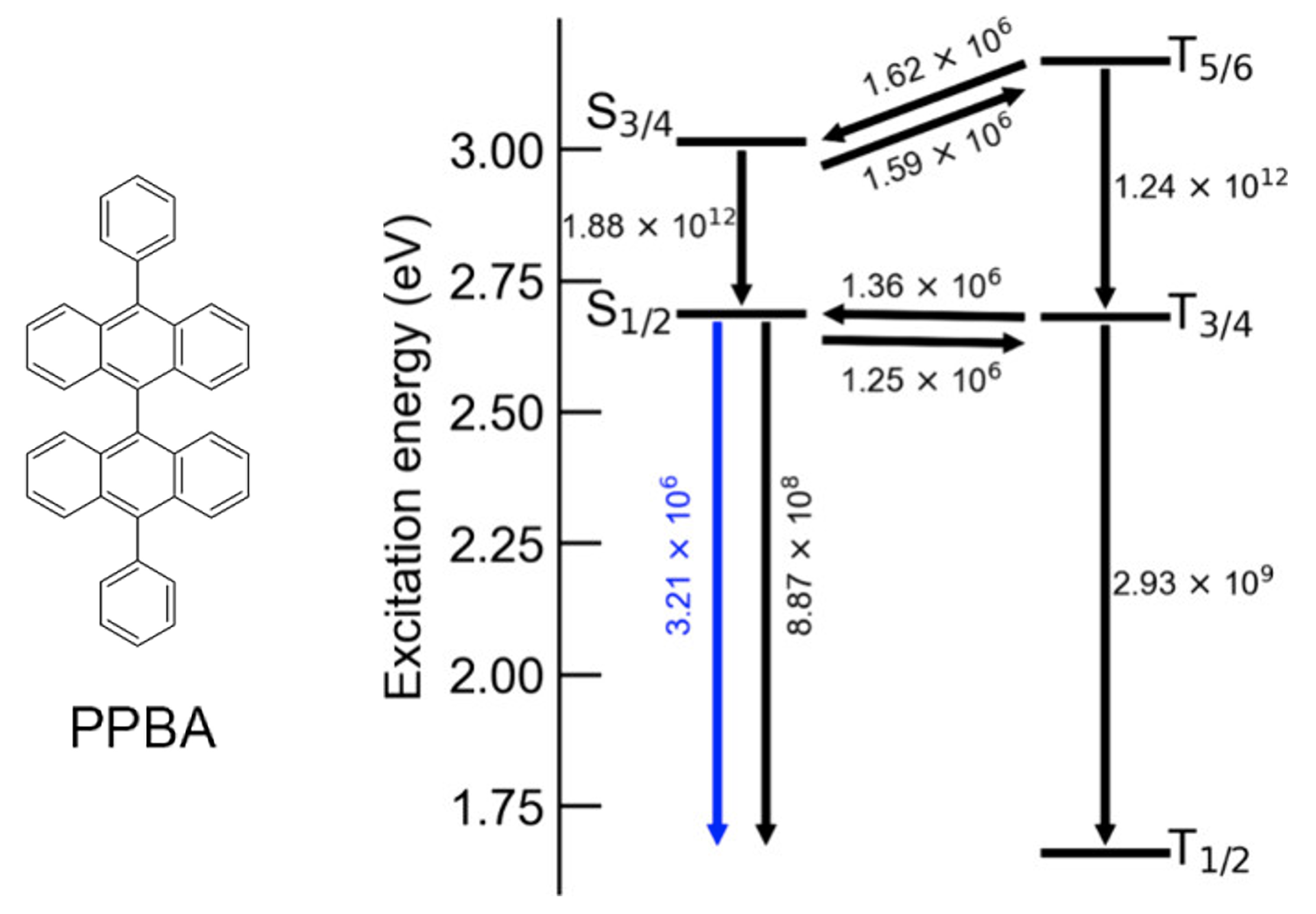}
    \caption{The structure of PPBA (left) and its FGR-based kinetic model (right). The black arrows represent non-radiative transitions with FGR--based rates, while the blue arrow shows the emission with an experimentally determined rate. Rates are given in the unit of s$^{-1}$. Reproduced with permission from J. Phys. Chem. A {\bf 129}, 3445 (2025) [Ref. \onlinecite{min-jpca129}]. Copyright  2025  American Chemical Society.}
    \label{fig:oled}
\end{figure}

Multiple-resonance (MR) emitters constitute another important class of OLED materials.\cite{MR}
In these systems, rigid molecular frameworks with alternating electron donating and
withdrawing units produce spatially separated but well-localized frontier orbitals,
leading to narrow emission spectra and moderate $\Delta E_{\rm{ST}}$. Despite relatively weak
spin-orbit coupling, RISC can be meaningfully efficient, often with vibronically assisted indirect coupling. This aspect has been rationalized using time-domain FGR expressions
combined with explicit vibrational analyses.\cite{kim-jacs-au-1}
Additional triplet harvesting mechanisms, such as the hot RISC (hRISC) pathway involving higher-lying triplet states, have also been analyzed within the framework of FGR.
For example, the diverse nonradiative transition rates of a hot exciton molecule, 10,10'-diphenyl-9,9'-bianthracene (PPBA), was studied with FGR to check the plausibility of singlet generation by hRISC (Fig.\ \ref{fig:oled}).\cite{min-jpca129} Similar kinetic modeling over multiple pathways toward determining the dominant path will also be possible for other systems.

Of course, in OLED materials, there may be multiple competing transitions that occur on comparable timescales, and evaluating FGR transition probabilities can provide inputs for constructing kinetic models to describe emission and other competing radiationless decays.\cite{ParkJW-FGR-ijms}

\subsubsection{Applications for photovoltaic materials}

Once formed, excitons in condensed phase may migrate through successive nonadiabatic transitions. In the weak coupling regime, FGR can provide a founding ground for reaching simple rate expressions that can explain migrations as hopping events without memories.\cite{Hsu-EET} A typical example will be the application toward organic photovoltaic (OPV) materials,\cite{or_pv,kippelen-ees2,lyons-ees5,sun-jpcc122,tong-jcp153,han-jctc16,hu-jpcb124,tinnin-jcp154,tinnin-jpcl13,schubert-jpcl14} where the exciton diffusion length is one of the key factors for deciding efficiency.\cite{exciton-diff} In OPV, exciton diffusion directly influences the overall success rate as a prefactor before reaching the stage of exciton dissociation at the donor--acceptor interfaces, and the consideration based on FGR framework may quantify how spectral overlap, energetic disorder, and related structural aspect influence the diffusion.
The ensuing charge separation also proceeds through nonadiabatic transitions from bound excitons to delocalized electron-hole pairs, and FGR can also quantify the rates of these processes when the coupling matrix element and the density of available acceptor states are known.\cite{solarcell} The nonradiative voltage loss through carrier recombination can also be studied with FGR.\cite{organic_HJ}
Of course, these formalisms share the same ground with the Marcus-type descriptions\cite{marcus-jcp24} of interfacial electron transfer.\cite{QME_CT_OPV}

At organic interfaces, the density of charge--transfer states increases with molecular packing and delocalization, which in turn facilitates charge separation from exciton states.\cite{BHJ-Barbatti-21,delocal-and-rate}
Similarly, morphological disorder originating from polymeric constructs broadens the density of state,\cite{polymerOPV}
and FGR should be able to capture the behaviors of such systems with ensemble averaged formulations. Indeed, simulations combining molecular dynamics and hopping kinetics based on FGR (or more precisely Marcus theory) have been performed to rationalize observed transport characteristics and to help propose design principles for improved bulk-heterojunction architectures.\cite{BHJ-jpcc-16,BHJ-pccp-21}

\section{Environmental effects}
\subsection{General aspects}
What are real effects of environments or media on FGR have long been debated,\cite{robinson-jcp37,bixon-jcp48} and accurate characterization of such effects is crucial for assessing true quantitative accuracy of FGR.  Environmental effects influence all components of FGR.  Considering the energy domain expression, Eq. (\ref{eq:kfgr-g}), the coupling Hamiltonian $\hat H_c$, the initial and final states $|\psi_j\rangle$ and $|\psi_f\rangle$, the initial distribution $p_j$, and the final energy density $\rho_f (E_j)$ all  depend on environments  in a nontrivial manner.   Considering the time domain expression, \rc{accurate account of environmental effects involves two aspects. One is how to determine the initial and final density operators $\hat \rho_j$ and $\hat \rho_f$ and the other is} how to represent the zeroth order and coupling Hamiltonian $\hat H_0$ and $\hat H_c$ so that accurate enough and efficient calculations can be conducted.  

Different fields of research have become successful in applying FGR by developing ways to recognize major environmental effects.  A prime example is the reorganization energy  introduced by  Marcus for ET,\cite{marcus-arpc15,marcus-bba811,marcus-jcp24,marcus-jcp26-1,marcus-jcp26-2,marcus-dfs29,marcus-jpc67,marcus-jcp43,marcus-arpc15,marcus-bba811,marcus-jcp81} which was  originally defined to account for the collective effects of solvent polarization with respect to the change in charge localization centers.  The concept of the  reorganization energy was then further extended to account for quantum molecular vibrations,\cite{jortner-jcp64} different time scales of environmental responses,\cite{hoffman-ica243,matyushov-jcp122,matyushov-jcp139} and has continued evolving further for more complex processes.\cite{adv-et,newton-tca110,blumberger-cr115,jang-rmp90}  

Some effects of environments can be ignored or greatly simplified for the purpose of determining the rate.   A well known example is the homogeneous line broadening of lineshape, which reflects various decoherence and relaxation processes involving environments.  It has been a long practice to model these simply as either Gaussian or Lorentzian function with empirically chosen width due to the  difficulty of accounting for all the complex interactions responsible for line broadening.  Good examples of simplification are modeling the  polarization response of environments in terms of macroscopic dielectric constants, and approximating the quantum environments in terms of  infinitely many harmonic oscillators. 

With experimental advances, the effects of environments ignored or simplified in earlier theories have become important in many cases and motivated further advances in theories\cite{dinpajooh-jcp146}  and computational methods. Even for calculating FGR rates in standard solid or liquid environments, satisfactory microscopic account of local dielectric responses at molecular scale, general anharmonic effects, time dependent fluctuations due to multiple sources in complex media remain challenging theoretical and computational issues.  In recent years, these issues have become more important with interest in rate processes in  novel or complex environments.  Some of these are described in the next subsections for more specific examples.

\subsection{Plasmonic systems}

Plasmons\cite{bohm-pr82,pines-pr85,bohm-pr92,pines-pr92,pines-rmp28} represent collective and delocalized states of electrons oscillating around the nuclear lattice in solid  and have been subject to extensive research in chemical physics, particularly in relation to surface enhanced Raman scattering.\cite{gersten-jcp73,weitz-jcp78,kelly-jpcb107,willets-arpc58}  Plasmons at the simplest level can be viewed as frequency dependent Drude-like oscillators and the enhancement effects of surface plasmons on the spectroscopy, charge transfer, and EET processes  that occur on the surface of bulk and nanoscale metals and semiconductors  are well known.  For these,  FGR has been frequently used\cite{gersten-cpl104,hua-jcp83,kelly-jpcb107,govorov-prb76,wang-jcp151,ding-jpcc122} by accounting for their effects on modification of coupling Hamiltonians, Franck-Condon overlaps, and the final density of states. 

\begin{figure}
    \centering
    \includegraphics[width=\linewidth]{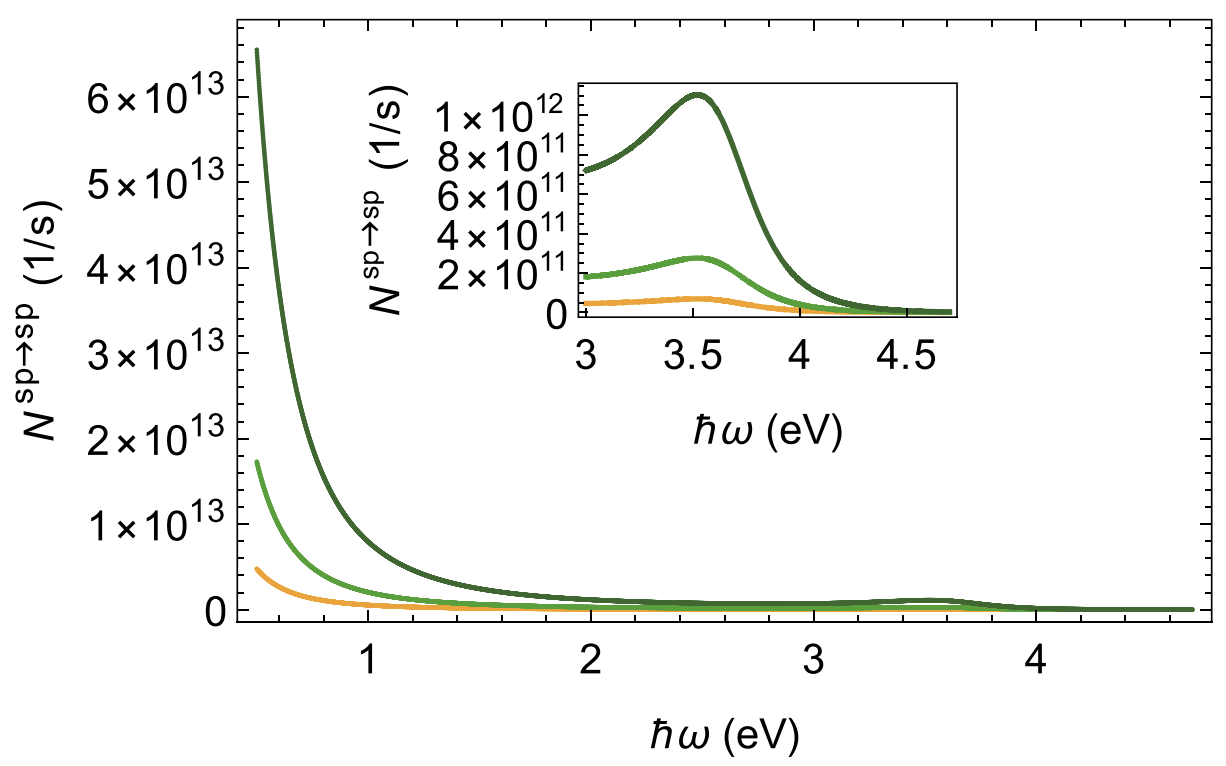}
    \caption{FGR-based plasmonic hot carrier generation rates as functions of incident photon energies, due to sp-band to sp-band transition of silver nano-particles with three different radii. Reproduced from [L. R. Castellanos, J. M. Kahk, O. Hess, and J. Lischner, J. Chem. Phys. {\bf 152}, 104111 (2020)], Ref. \onlinecite{castellanos-jcp152}, with the permission of AIP Publishing. }
    \label{fig:hot_carrier}
\end{figure}

The dynamics within and across  plasmon systems themselves have also been important subjects of research that employed FGR.  For example, FGR can be used to quantify the generation and relaxation rates of hot carriers resulting from localized surface plasmon decay.\cite{HotCarrierTheory-21a,HotCarrierTheory-21b} 
Figure \ref{fig:hot_carrier} shows the FGR-based generation rates of plasmonic hot carriers in silver nano-particles, which demonstrate significant dependence of  hot carrier generation on the radius of the particle.\cite{castellanos-jcp152}  Controlling such dependence is crucial for optimizing quantum yields and improving the efficiency of photodetectors, plasmonic catalysts, and energy harvesting devices. More specifically, the ability to accurately model carrier dynamics will allow enhancing the performance of plasmonic nanomaterials in energy conversion applications. By capturing complex interactions and various pathways, these models help optimize the efficiency of plasmonic catalysts for applications toward environmental remediation and fuel production.

Despite the success of many theories and computational methods involving plasmons, accurate microscopic calculation of plasmon systems as environments remains challenging.  This is particularly an important issue for more satisfactory understanding of plasmon enhancement in the presence of molecular systems.  The interaction between these molecular systems and plasmon systems will alter spatio-temporal  characteristics of both sides.  Further advances in theories and computational methods are needed to address these issues.  

\subsection{Magnetic systems}

Magnetic materials and related spin dynamics have long been important subjects of chemical physics.  While earlier studies were linked to nuclear magnetic resonance and electron spin resonance, many of recent studies have implications for the development of quantum information processing devices broadly.   Recent surge of interest in chirality induced spin selection (CISS),\cite{bloom-cr124} for which there is no clear theoretical consensus yet, is also motivating new theoretical advances intended to describe coupling between spins and electron-nuclear dynamics at detailed molecular level.   

Interactions involving spins are weak relative to other interactions in molecules and justify applications of  FGR as far as incoherent dynamics are concerned.    Indeed, FGR has  been routinely used for the description of spin transitions in magnetic environments. For example, FGR provided microscopic description of spin flip processes in a rare-earth ion--doped crystal systems.  For these processes, magnetic dipole-dipole interactions around ionic dopants serve as the coupling Hamiltonian and offered the foundation for  understanding  macroscopic and collective magnetic behaviors.\cite{spin-flip-flop} 

Driven magnetization and spin transitions can also be modeled by FGR to some extent. For example,  it was shown that FGR can be used to model optically induced spin flip processes on two-dimensional nonmagnetic surfaces.\cite{spinflip} FGR can be used for characterization of optically induced spin flip transition rates and/or subsequent relaxation rates, and can be used to assess the influence of local structural variations on spin flips, particularly in low-temperature environments where quantum effects are pronounced.

FGR can also offer key theoretical understanding of transitions in spin cross-over materials.  For example, FGR was used for  electron transport through spin-crossover molecular junctions.\cite{electrontransport} In these systems, electronic transport and spin state dynamics are closely linked, influencing the performance of the resulting spintronic devices. Application of FGR thus allows calculating the conductance of spin-crossover molecules, which in turn helps optimizing the functionality at molecular level.

\subsection{Photonic and optoelectronic systems}

As described  in the context of spontaneous emission in  Sec. V.A.1, FGR provides a simple but clear description of the Purcell effect,\cite{Purcelleff} which represents the modification of emission rates depending on the electromagnetic environment of an emitter, \rc{and other effects that are important in engineered photonic environments.\cite{david-rpp75}   Such effects can also be} prominent in chiral media, where the interaction between electromagnetic fields and chiral molecules significantly alters emission rates.\cite{PurcelleffChiral} FGR also allows straightforward understanding of  emission enhancement and suppression in those chiral systems,\cite{purcell19} unless its underlying assumptions break down due to very strong phonon coupling.\cite{breakdownFGR15}

Controlling light emission from quantum dots is an important issue in relation to the development of novel optoelectronic devices.\cite{QLED}  In particular, how to minimize losses of quantum-yield  arising from charge recombination dynamics in quantum-confined structures has important practical implications, for which FGR offers a key guideline.\cite{QD-loss} The presence of functional groups and dopants in quantum dots modifies the electronic states and their coupling to vibrational modes.  FGR can be used to elucidate how these factors dictate recombination rates, and thus can help optimizing the quantum efficiency of quantum dot based systems.

In wide-bandgap materials, such as $\beta$-${\rm Ga_2O_3}$, the thermalization of radiation-induced carriers can be analyzed using FGR,\cite{Ga2O3} allowing to model carrier-phonon scattering and to optimize the performance of materials for radiation detection and its electronic applications.  More generally, even in quantum many-body systems,  FGR can be applied to understand similar pre-thermalization and thermalization processes, providing insights into how quantum systems evolve toward equilibrium.

Spectroscopies and EET processes in engineered photonic environments have long been subjects of experimental and theoretical studies.\cite{andrew-science290,wubs-njp18}  Recent advances in the creation of cavity systems and the possibility of significant effects of polaritons, the matter-radiation hybrid states, has led to numerous experimental and theoretical studies.  FGR level theories\cite{du-cs9,wang-jcp151,wu-jpcl9,saller-jpcl13,saller-jpcc127} that account for proper initial and final states and coupling mechanisms developed for these systems offer key insights into novel applications of these systems.  

\subsection{Biological and disordered organic environments}
FGR has long been used for charge, proton, and energy transfer processes in biological and disordered organic environments.  These environments are complex in the sense that networks of various length scales, environmental dynamics of broad time and energy scales, and significant disorder and fluctuations exist.  Therefore, accurate and consistent calculation of all the factors contributing to FGR is necessary for reliable description.  It is often the case that smallness of coupling Hamiltonian is the result of interference of many coupling pathways\cite{adv-et,beratan-science258,balabin-science290,stuchebrukhov-jcp104,stuchebrukhov-jcp105} or renormalization by environmental responses.\cite{jang-rmp90} The estimates for reorganization energy can be substantially different depending on the types of spectroscopic measurements and calculation.  It is often inevitable to introduce phenomenological Gaussian distribution\cite{jang-rmp90} for quantitative modeling of experimental data.  Nonetheless, significant progress has been made through combination of calculations based on coarse-grained models, molecular dynamics, mixed quantum-classical calculation, and validation through comprehensive experimental data.  These include enzyme catalysis,\cite{hammes-schiffer-cr110} energy and charge transfer processes in natural photosynthetic systems,\cite{jang-rmp90,PhotoSyn} organic light emitting diodes,\cite{burroughes-nature347,kohler-am14,sasaki-nc12,zou-mcf4} and bulk hetero junction photovoltaic devices,\cite{or_pv,kippelen-ees2,lyons-ees5,jailaubekov-nm12,lee-jpcl5} as have been mentioned in Sec. V. 

There have also been interesting proposals that suggest novel quantum effects in biological systems based on FGR level theories.  For example, the vibrational theory of olfaction (VTO)\cite{brookes-prl98} is a standard application of FGR, and magneto-reception involving transitions of triplet states\cite{magnetoreception} was developed based on FGR.  However, the success and the reliability of these theories hinge on the validity of the models even though the application of FGR itself is correct.  For example,  VTO was deemed implausible because quantum treatments of protein environments can easily screen the olfactory sensitivity proposed by VTO.\cite{block-pnas112-1} There are also ample experimental evidences against VTO.\cite{block-pnas112-1} Regarding theories of magnetoreception, the issues are more complex.  While the theory seems to consider the environmental effects to a quite reasonable level, there is still a significant gap between the underlying model and the actual biological system.    Thus, further computational and experimental data are needed for the validation of this theory as realistic representation of the actual biological process.

\section{Concluding Remarks}
FGR is the simplest and oldest quantum theory of transition and has served as the fundamental basis for theoretical understanding of many spectroscopic signals and quantum transfer/transport processes.  Although it is a textbook level theory, its refinement for practical applications in chemical physics is not. Ingenious theoretical and computational advances have been made to identify and calculate appropriate zeroth order and coupling Hamiltonians, which have led to foundational theories for spectroscopy and quantum transfer/transport processes. 

The conceptual clarity of FGR has enabled numerous fruitful experiment-theory collaborations.  The generality of FGR has made it possible to develop broadly applicable and quantitative theories, in combination with advanced computational methods.   On the other hand, the simplicity of assumptions underlying FGR has left room for further generalizations  for situations that go beyond its original assumptions.  Examples are nonequilibrium and/or non-Markovian generalizations\cite{coalson-jcp101,evans-jcp104,jang-cp275,jang-jcp122,jang-jcp127,jang-prl92,jang-jpcb110,jang-wires3,jang-prl113,sun-jctc12,sun-jcp144,sun-jcp145} and  Floquet-FGR theory\cite{kohler-pre58,ikeda-prb104} for periodically driven Hamiltonians. 

With recent advances in computational methods and experimental precision,  quantitative assessment of FGR-level theories has become possible more than ever.  Many of such  tests have shown surprising accuracy, when appropriate FGR-level theories are used in the proper context.   Future advances in electronic structure calculation methods, coarse-graining approaches, and machine learning approaches will improve these aspects of FGR even further. 
Even if the accuracy  of a given FGR-level theory may not be satisfactory, it can  still provide an important reference for the nature of  improvement gained through an advanced quantum dynamics calculation method.  

In \rc{making} FGR-level theories and their generalizations more versatile and accurate, there remain important issues to account for.   These include the effects of non-Condon couplings, anharmonicity, and time dependent fluctuations.  Although significant theoretical and computational advances have been made to address these factors, they are not yet general enough for routine and reliable applications.   As a result, applications of existing theories  are often accompanied by additional assumptions and approximations that may introduce new sources of error.  Nonetheless, FGR offers the most general and flexible framework for  developing theories addressing such issues.   In addition, development of an effective FGR-level theory is expected to be the first step in emerging fields such as CISS.\cite{bloom-cr124}  Thus, it is likely that significant advances in formulations and computational methods based on FGR will continue in the foreseeable future.

\appendix 
\section{Steady state rate from the first order time dependent perturbation theory \label{app-a}}
This section summarizes the conventional derivation of FGR.  First, the perturbation Hamiltonian is assumed to be a time dependent $\hat H_c(t)$, which is not necessarily Hermitian.   In addition, it is assumed that $|\psi_f\rangle$ is not necessarily an eigenstate of $\hat H_0$, whereas $|\psi_j\rangle$ is an eigenestate and $\langle \psi_j|\psi_f\rangle=0$.  
The first order approximation for the transition amplitude for this case is the same as Eq. (\ref{eq:tr_am-1st}) except that $\hat H_c$ is replaced with a time dependent $\hat H_c(t)$.  Thus, the time dependent probability for the transition from $|\psi_j\rangle$ to $|\psi_f\rangle$, up to the second order of  $\hat H_c(t)$, is 
\ben
 &&p_{_{j\rightarrow f}}^{(2)}(t_s)\equiv |c_{jf}^{(1)}(t_s)|^2 \nonumber \\
 &&=\frac{1}{\hbar^2} \int_0^{t_s} dt \int_0^{t_s} dt' \langle \psi_f| e^{-i\hat H_0 (t_s-t)/\hbar}\hat H_c(t) e^{-i\hat H_0 t/\hbar}|\psi_j\rangle  \nonumber  \\
 &&\hspace{0.8in}\times \langle \psi_j|e^{i\hat H_0 t'/\hbar} \hat H_c^\dagger (t')e^{i\hat H_0 (t_s-t')/\hbar} |\psi_f \rangle .
 \een
We can define a time dependent rate by taking the derivative of $p_{_{j\rightarrow f}}^{(2)}(t_s)$ as follows:
\ben
 &&k_{_{j\rightarrow f}}(t_s)\equiv \frac{d}{d t_s} p_{_{j\rightarrow f}}^{(2)}(t_s)  \nonumber \\
 &&=\frac{2}{\hbar^2} {\rm Re} \int_0^{t_s} dt  \langle \psi_f| e^{-i\hat H_0 (t_s-t)/\hbar}\hat H_c(t) e^{-i\hat H_0 t/\hbar}|\psi_j\rangle  \nonumber  \\
 &&\hspace{1.2in}\times \langle \psi_j|e^{i\hat H_0 t_s/\hbar} \hat H_c^\dagger (t_s) |\psi_f \rangle . 
 \een
Since $|\psi_j\rangle$ is an eigenstate of $\hat H_0$, the above expression can be written as 
\ben
 &&k_{_{j\rightarrow f}}(t_s)=\frac{2}{\hbar^2} {\rm Re} \int_0^{t_s} dt  \langle \psi_f| e^{-i\hat H_0 (t_s-t)/\hbar}\hat H_c(t) e^{i\hat H_0 (t_s-t)/\hbar}  \nonumber  \\
 &&\hspace{1.2in}\times |\psi_j\rangle \langle \psi_j|\hat H_c^\dagger (t_s) |\psi_f \rangle . 
 \een
Replacing the integration variable $t$ with $t_s-t$, and introducing $\hat \rho_j=|\psi_j\rangle\langle \psi_j|$, the above expression can be written as 
\ben
 &&k_{_{j\rightarrow f}}(t_s)\nonumber \\
 &&=\frac{2}{\hbar^2} {\rm Re} \int_0^{t_s} dt \langle \psi_f| e^{-i\hat H_0 t/\hbar}\hat H_c(t_s-t) \hat \rho_j  e^{i\hat H_0 t/\hbar} \hat H_c^\dagger (t_s) |\psi_f\rangle .\nonumber \\ \label{eq:kif-time}
 \een

Let us assume that there exists a long enough value of $t_s$ such that the condition ${\bf C1}$ in the main text is satisfied.  Let us also assume that $|\psi_f\rangle$ is an eigenstate of $\hat H_0$. Then, given that the integral in the above equation is well defined in the limit of $t_s\rightarrow \infty$, Eq. (\ref{eq:kif-time}) approaches well-known FGR rate expressions. 
 For example, for time independent and Hermitian $\hat H_c$ and $|\psi_f\rangle$, it reduces to 
  \ben
 &&k_{_{{\rm F},j\rightarrow f}}=\frac{2}{\hbar^2} {\rm Re} \int_0^{\infty} dt  e^{i(E_j-E_f)t/\hbar} |\langle \psi_f| \hat H_c |\psi_j\rangle|^2 ,
 \een
 which is equivalent to Eq. (\ref{eq:kfgr-0}) when Eq. (\ref{eq:delta}) is used for the Dirac-delta function.  
 For the case where $\langle \psi_f|\hat H_c(t)|\psi_j\rangle=\langle \psi_f|\hat h_c |\psi_j\rangle e^{-i\omega t} $, Eq. (\ref{eq:kif-time}) reduces to 
  \ben
 &&k_{_{{\rm F},j\rightarrow f}}\nonumber \\
 &&=\frac{2}{\hbar^2} {\rm Re} \int_0^{\infty} dt  e^{i[\omega  +(E_j-E_f)/\hbar]t } |\langle \psi_f| \hat h_c|\psi_j \rangle|^2   , \nonumber \\  \label{eq:ks-eigen}
 \een
 which becomes the standard FGR rate expression for a spectroscopic transition when Eq. (\ref{eq:delta}) is used for the Dirac-delta function.
  
\section{Derivation of FGR rate expression for states interacting with bath \label{app-b}}
 Assume that we are interested in a system that can be completely represented by a set of orthonormal states, each denote as $|\varphi_j\rangle$.  Thus, the identity operator in the Hilbert space of the system is assumed to be $\hat 1_s=\sum_j|\varphi_j\rangle\langle \varphi_j|$.     The remainder of the total system is denoted as bath.  Then, assume that the total Hamiltonian $\hat H(t)=\hat H_0+\hat H_c(t)$, where 
\ben
&&\hat H_0=\sum_j ({\mathcal E}_j +\hat B_j)|\varphi_j\rangle\langle \varphi_j|+\hat 1_s\otimes \hat H_b,  \label{eq:h0t-gen} \\
&&\hat H_c(t)=\sum_j \sum_{k\neq j}\hat J_{jk}(t)|\varphi_j\rangle\langle \varphi_k|  .\label{eq:hct-gen}
\een
In the above expressions, $\hat B_j$ is a time independent bath operator, $\hat H_b$ is the bath Hamiltonian governing the dynamics in the absence of system, and   $\hat J_{jk}(t)$ is a coupling term that can be dependent on both time and bath degrees of freedom.  Then, given that $\hat \rho_j=|\varphi_j\rangle\langle  \varphi_j|\hat \rho_{b,j}$ and $\hat \rho_f=|\varphi_f\rangle\langle \varphi_f|\hat 1_b$, Eq. (\ref{eq:kif-time-1}) can be expressed as 
\ben
&&k_{_{j\rightarrow f}}(t_s)=\frac{2}{\hbar^2} {\rm Re} \int_0^{t_s} d t e^{i({\mathcal E}_j-{\mathcal E}_f)t/\hbar} \nonumber \\
&&\hspace{.2in}\times   {\rm Tr}_b\left\{ \hat J_{jf}^\dagger(t_s) e^{-i (\hat B_f+\hat H_b) t/\hbar}\hat J_{jf} (t_s-t) \hat \rho_{b,j} e^{i t ( \hat B_j+\hat H_b)/\hbar } \right\} ,\nonumber \\  \label{eq:kt-fi}
\een 
where ${\rm Tr}_b$ is the trace over the bath. 
Note that  $\hat B_j+\hat H_b$ commute with $\hat \rho_{b,j}$ by assumption. 

Now let us consider a simple and generic case of time independent  $\hat H_0$ and $\hat H_c$ given by
\ben
\hat H_0&=&\sum_j  ({\mathcal E}_j +\hat B_j)|\varphi_i\rangle\langle \varphi_j|+\hat H_b \nonumber \\
\hat H_c&=&\sum_j \sum_{k\neq j}J_{jk}|\varphi_j\rangle\langle \varphi_k|  \label{eq:ht-gen-1}
\een
with
\ben
&&\hat B_j=\sum_n \hbar\omega_n g_{nj}(\hat b_n+\hat b_n^\dagger) , \label{eq:bj-lin}\\
&&\hat H_b=\sum_n \hbar\omega_n \left (\hat b_n^\dagger \hat b_n+\frac{1}{2}\right) . \label{eq:hb-har}
\een
In the above expression, $\omega_n$, $\hat b_n$, and $\hat b_n^\dagger$ are frequency, lowering operator, and raising operator of the $n$th harmonic oscillator mode of the 
bath, whose coupling strength to state $|j\rangle$ is given by $g_{nj}$.
Then, assuming $\hat \rho_{b,j}=\hat \rho_{b,j}^{eq}$, Eq. (\ref{eq:kt-fi}) for the present case can be calculated explicitly\cite{jang-exciton} and expressed as
\be
k_{_{j\rightarrow f}}(t_s)=\frac{2|J_{jf}|^2}{\hbar^2} {\rm Re} \left [ \int_0^{t_s} d t e^{i(\tilde {\mathcal E}_j-\tilde {\mathcal E}_f) t/\hbar -{\mathcal C}_{jf}(t) } \right ]  , \label{eq:kt-2}
\ee
where  
\ben 
&&\tilde {\mathcal E}_j={\mathcal E}_j- \sum_n\hbar\omega_n g_{nj}^2  \label{eq:rel-eng}\\
&&{\mathcal C}_{jf}(t)=\sum_n \delta g_{n,jf}^2 \Big \{ \coth \left (\frac{\beta\hbar\omega_n}{2}\right) \left (1-\cos (\omega_n t)\right)   \nonumber \\
&&\hspace{.5in}   +i\sin (\omega_n t)  \Big \} , \label{eq:ct}
\een
with  $\delta g_{n,jf}=g_{nj}-g_{nf}$.  Employing the following bath spectral density:
\be
{\mathcal J}_{jf}(\omega)=\pi\hbar \sum_n \delta g_{n,jf}^2 \delta (\omega-\omega_n) \omega_n^2 , \label{eq:bath-sp}
\ee
 Eq. (\ref{eq:ct}) can also be expressed as 
 \ben
{\mathcal C}_{jf}(t)&=&\frac{1}{\pi\hbar}\int_0^\infty d\omega \frac{{\mathcal J}_{jf}(\omega)}{\omega^2} \Big \{ \coth \left (\frac{\beta\hbar\omega}{2}\right) \left (1-\cos (\omega t)\right)   \nonumber \\
&&\hspace{.5in}   +i\sin (\omega t)  \Big \}  \nonumber \\
&&={\mathcal C}_{R,jf}(t)+i{\mathcal C}_{I,jf}(t) , \label{eq:ct-2}
\een
where ${\mathcal C}_{R,jf}(t)$ and ${\mathcal C}_{I,jf}(t)$ are real and imaginary parts.

\section{Stationary phase approximation for Eq. (\ref{eq:kf-3}) \label{spa}}
Extending the integration in Eq. (\ref{eq:kf-3}) to the complex domain,  it can be expressed as 
 \be
k_{_{{\rm F},j\rightarrow f}}=\frac{|J_{jf}|^2}{\hbar^2} \int_C d z\ e^{ f(z) }   , \label{eq:kf-2-st0}
\ee
where $C$ denotes appropriately deformed contour in the complex domain and  $f(z)=-i \Delta G_{jf}z/\hbar -{\mathcal C}_{jf}(z)$.    In this expression, ${\mathcal C}_{jf}(z)$ is defined by Eq. (\ref{eq:ct}) or (\ref{eq:ct-2}), but using $z$ as its argument instead, with an obvious definition of $\Delta G_{jf}$.
For the present section, it is convenient to rewrite  ${\mathcal C}_{jf}(z)$ as follows: 
\ben
&&C_{jf}(z)=\sum_n \delta g_{n,jf}^2 \frac{\cosh \left (\frac{\beta\hbar\omega_n}{2}\right) -\cosh \left (\frac{\beta\hbar\omega_n}{2}-i\omega_n z\right )}{\sinh \left (\frac{\beta\hbar\omega_n}{2}\right) }\nonumber \\
&&=\frac{1}{\pi\hbar}\int_0^{\infty}  d\omega \frac{{\mathcal J}_{jf}(\omega)}{\omega^2} \frac{\cosh \left (\frac{\beta\hbar\omega}{2}\right) -\cosh \left (\frac{\beta\hbar\omega}{2}-i\omega z\right )}{\sinh \left (\frac{\beta\hbar\omega}{2}\right) } .\nonumber \\ \label{eq:cz-exp}
\een
Let us assume that there exists a stationary point $z_s$ such that $f'(z_s)=0$.  Then, $f(z)$ can be approximated around $z_s$ as 
\be
f(z)\approx f(z_s)+\frac{1}{2} f''(z_s)(z-z_s)^2 \label{eq:f-2nd}
 \ee
Let us also assume that $f(z_s)$ is real and $f''(z_s)$ is real and negative.   Then, choosing a  contour $C$  such that it passes through $z_s$ while being parallel to the real axis and conducting the gaussian integration under the approximation of Eq. (\ref{eq:f-2nd}), we obtain
 \be
k_{_{{\rm F},j\rightarrow f}}\approx \frac{|J_{jf}|^2}{\hbar^2} \sqrt{\frac{2\pi}{|f''(z_s)|}} e^{ f(z_s) }   . \label{eq:kf-2-st}
\ee

Now let us consider $f(z)$ in more detail to identify $z_s$ that meets the condition validating Eq. (\ref{eq:kf-2-st}).
Taking the derivative of $f(z)$, using the expression for $C_{jf}(z)$ given by Eq. (\ref{eq:cz-exp}), we obtain the following equation for $z_s$:
\ben
\frac{\Delta G_{jf}}{\hbar}&=&-\sum_n \frac{\delta g_{n,jf}^2 \omega_n}{\sinh (\frac{\beta\hbar\omega_n}{2})} \sinh \left (\frac{\beta\hbar\omega_n}{2}-i\omega_n z_s\right)  \nonumber \\
&=&-\frac{1}{\pi\hbar}\int_0^\infty d\omega \frac{{\mathcal J}_{jf}(\omega)}{\omega \sinh (\frac{\beta\hbar\omega}{2}) }   \sinh \left (\frac{\beta\hbar\omega}{2}-i\omega z_s\right) .\nonumber \\ \label{eq:sta-eqn}
\een
On the other hand, the second derivative is 
\ben
f''(z_s)&=&-C_{jf}''(z_s)=-\sum_n \delta g_{n,jf}^2 \omega_n^2  \frac{\cosh \left (\frac{\beta\hbar\omega_n}{2}-i\omega_n z_s\right )}{\sinh \left (\frac{\beta\hbar\omega_n}{2}\right) }\nonumber \\
&=&-\frac{1}{\pi\hbar}\int_0^{\infty}  d\omega {\mathcal J}_{jf}(\omega) \frac{\cosh \left (\frac{\beta\hbar\omega}{2}-i\omega z_s\right )}{\sinh \left (\frac{\beta\hbar\omega}{2}\right) } . \label{eq:f-2nd-der}
\een

  Noting the definition of $\lambda$ given by Eq. (\ref{eq:lambda-def}), it is easy to identify a solution of Eq. (\ref{eq:sta-eqn}) for three particular values of $\Delta G_{jf}$, which are $z_s^*=0$, $-i\beta\hbar/2$, and $-i\beta\hbar$ respectively for $\Delta G_{jf} =-\lambda$, $0$, and $\lambda$.  For these values, $f''(z_s^*)$ given by Eq. (\ref{eq:f-2nd-der}) is real and negative. 
 Making a linear expansion of $z_s$ around these points in the integrand of Eq. (\ref{eq:sta-eqn}) then leads to the following three approximate solutions around the three values of $\Delta G_{jf}$ as summarized below. 
\be
z_s=\left \{ \begin{array}{ll} -i\beta\hbar (\Delta G_{jf}+\lambda)/(2\lambda_{qc}), &\mbox{ for }\Delta G_{jf} \sim -\lambda\\  -i\beta\hbar (\Delta G_{jf}+\lambda_{qs})/(2\lambda_{qs}),   &\mbox{ for }\  \Delta G_{jf} \sim 0\\  -i\beta\hbar (\Delta G_{jf}+2\lambda_{qc}-\lambda )/(2\lambda_{qc}) ,  &\mbox{ for }\Delta G_{jf} \sim \lambda \end{array}\right .
\ee
For these values of $z_s$, $f(z_s)$ is \rc{real-valued} and $f''(z_s)<0$ for small enough $\Delta G_{jf}$ within an order $\lambda$.  Equations (\ref{eq:kf-2-sca-1}), (\ref{eq:kf-2-spa-1}), and (\ref{eq:kf-2-spa-2}) respectively correspond to Eq. (\ref{eq:kf-2-st}) using $f(z_s)$ expanded up to second order of  $z_s-z_s^*$ while using $f''(z_s^*)$ for each of the three values of $\Delta G_{jf}=-\lambda$, $0$, and $\lambda$.     

 \section{Projection operator formulation of master equation and transition rates\label{app-c}}
 With the definition of system Hilbert space spanned by $|\varphi_j\rangle$\rc{'}s and the remainder as the bath, as described in Sec. \ref{app-b}, it is possible to decompose any general Hamiltonian defined in the direct product space of the system and the bath into the zeroth and coupling terms.  For a general time dependent total Hamiltonian $\hat H(t)$, let us define 
\be
\hat H_{jk}(t)=\langle \varphi_j|\hat H (t) |\varphi_k\rangle ,
\ee
which remains as an operator in the Hilbert space of the bath.  
Then, we can define
\be
\hat H_0(t)=\sum_j |\varphi_j\rangle \hat H_{jj}(t) \langle  \varphi_j| , 
\ee
\be 
\hat H_c(t)=\sum_{j}\sum_{k \neq j} |\varphi_j\rangle \hat H_{jk}(t) \langle \varphi_k|  . \label{eq:hct-def}
 \ee
 Note that $\hat H(t)=\hat H_0(t)+\hat H_c(t)$. 
 
 Given the total density operator $\hat \rho(t)$ defined in the direct product space of the system and the bath, the probability for the state to be in the state $|\varphi_j\rangle$ is given by
 \be
 p_j(t)={\rm Tr}_b \left\{ \langle \varphi_j|\hat \rho(t)|\varphi_j\rangle \right\} ={\rm Tr} \left \{ |\varphi_k\rangle\langle \varphi_k| \hat \rho(t) \right\} ,
 \ee   
 where ${\rm Tr}_b$ is the trace over the bath and ${\rm Tr}$ is the trace over the total system and bath.  The time evolution of the total density operator $\rho(t)$ is governed by 
\ben
\frac{d}{dt}\hat \rho(t)&=&-i\left ({\mathcal L}_0(t)+{\mathcal L}_c(t)\right) \hat \rho (t) \nonumber \\
&\equiv&-\frac{i}{\hbar} \left [ \hat H_0(t)+ \hat H_c(t),\hat \rho(t)\right ] , \label{eq:ql-total}
\een
where ${\mathcal L}$ represents quantum Liouville operator and is defined by the second equality.
In the interaction picture with respect to $\hat H_0(t)$, the density operator becomes
\be
\hat \rho_I(t) =\hat U_0^\dagger (t) \hat \rho(t) \hat U_0(t)  ,
\ee
where 
\be 
\hat U_0(t)= e_{(+)}^{-i\int_0^t d\tau \hat H_0(\tau)/\hbar}=\sum_j|\varphi_j\rangle\langle \varphi_j|\hat U_{0,j}(t) ,
\ee   
with $\hat U_{0,j}(t)=e_{(+)}^{-i\int_0^t d\tau \hat H_{jj}(\tau)/\hbar}$.
Then, Eq. (\ref{eq:ql-total}) can be transformed into the following time evolution equation for $\hat \rho_I(t)$:
\be
\frac{d}{dt} \hat \rho_I(t)=-i{\mathcal L}_{c,I}(t)\hat \rho_I(t) \equiv -\frac{i}{\hbar}\left [\hat H_{c,I}(t),\hat \rho_I(t)\right ] , \label{eq:drho-lcit}
\ee
where 
\ben
\hat H_{c,I}(t)&=&\hat U_0^\dagger(t)\hat H_c(t)\hat U_0(t)  \nonumber \\
 &=& \sum_j\sum_{k\neq j} |\varphi_j\rangle \hat U_{0,j}(t) \hat H_{jk}(t) \hat U_{0,k}^\dagger (t) \langle \varphi_k| \label{eq:hcit}
\een
Let us then introduce the following projection super-operator:
\be
{\mathcal P}(\cdot) =\sum_j |\varphi_j\rangle\langle \varphi_j|\hat \rho_{b,j}{\rm Tr}\left \{|\varphi_j\rangle\langle \varphi_j |(\cdot)\right\} , 
\ee
where $(\cdot)$ represents an arbitrary operator  and $\hat \rho_{b,j}$ is an appropriate bath density operator that will be specified later in this section.
Note that 
\be
{\mathcal P}(\hat \rho (t))=\sum_j p_j(t) |\varphi_j\rangle\langle \varphi_j|\hat \rho_{b,j} .
\ee
In addition, it is straightforward to show that ${\mathcal P}{\mathcal L}_{c,I}(t){\mathcal P}=0$. With this property, application of the general identity and formal solution for the projection operator results in the following formally exact time evolution equation:
\begin{widetext}
\ben
\frac{d}{dt}{\mathcal P} \hat \rho_I(t)&=&-{\mathcal P}{\mathcal L}_{c,I}(t)(1+i\hat \Gamma_{c,I}(t))^{-1}\int_0^t d\tau e_{(+)}^{-i\int_\tau^t d\tau'  {\mathcal Q} {\mathcal L}_{c,I} (\tau')} {\mathcal Q} {\mathcal L}_{c,I}(\tau) {\mathcal P} e_{(-)}^{i\int_\tau^t d\tau' {\mathcal L}_{c,I}(\tau')}{\mathcal P} \hat \rho_{I}(t) \nonumber \\
&&-i {\mathcal P}{\mathcal L}_{c,I}(t)(1+i\hat \Gamma_{c,I}(t))^{-1}  e_{(+)}^{-i\int_0^t d\tau  {\mathcal Q} {\mathcal L}_{c,I} (\tau)}{\mathcal Q}\hat \rho_I (0) , \label{eq:proj-formal}
\een
 \end{widetext}
 where ${\mathcal Q}=1-{\mathcal P}$ and 
 \be
 \hat \Gamma_{c,I}(t)=\int_0^t d\tau e_{(+)}^{-i\int_0^t d\tau' {\mathcal Q} {\mathcal L}_{c,I}(\tau')} {\mathcal Q} {\mathcal L}_{c,I}(\tau) {\mathcal P}  e_{(-)}^{i\int_0^t d\tau'  {\mathcal L}_{c,I}(\tau')}
 \ee
Now assume that the initial condition of the total density operator is such that $\hat \rho(0)=|\varphi_{j_0}\rangle\langle \varphi_{j_0}|\hat \rho_{b,j_0}$ for a certain initial index $j_0$.  For this choice, ${\mathcal Q}\hat \rho_I (0)=0$ in Eq. (\ref{eq:proj-formal}).  Then, taking trace of Eq. (\ref{eq:proj-formal}), we obtain 
\be
\sum_j |\varphi_j\rangle\langle \varphi_j| \hat \rho_{b,j}\frac{d}{dt} p_j(t) =-\sum_{j,k} |\varphi_j\rangle\langle \varphi_j|\hat \rho_{b,j} {\mathcal R}_{jk} (t) p_k(t)  ,\label{eq:me-formal-0}
 \ee 
where
\begin{widetext}
\be
{\mathcal R}_{jk}(t)=\int_0^t d\tau \langle \varphi_j| {\rm Tr_b} \left \{ {\mathcal L}_{c,I}(t)(1+i\hat \Gamma_{c,I}(t))^{-1}  e_{(+)}^{-i\int_\tau^t d\tau'  {\mathcal Q} {\mathcal L}_{c,I} (\tau')} {\mathcal Q} {\mathcal L}_{c,I}(\tau) {\mathcal P} e_{(-)}^{i\int_\tau^t d\tau' {\mathcal L}_{c,I}(\tau')} |\varphi_k\rangle\langle \varphi_k|\hat \rho_{b,k} \right\} |\varphi_j\rangle   . 
\ee
\end{widetext} 
Taking trace of Eq. (\ref{eq:me-formal-0})  over the bath and considering the component of $|\varphi_j\rangle\langle \varphi_j|$, we obtain the following master equation:
\be
\frac{d}{dt} p_j(t) =-\sum_{k}  {\mathcal R}_{jk} (t) p_k(t)  . \label{eq:me-formal-1}
 \ee 
When considered up to the second order of $\hat H_{c,I}(t)$, ${\mathcal R}_{jk}(t)$ is approximated as
\be
{\mathcal R}_{jk}^{(2)}(t)=\int_0^t d\tau \langle \varphi_j| {\rm Tr_b} \left \{ {\mathcal L}_{c,I}(t)   {\mathcal L}_{c,I}(\tau) |\varphi_k\rangle\langle \varphi_k|\hat \rho_{b,k} \right\} |\varphi_j\rangle  . 
\ee
Employing Eqs. (\ref{eq:hct-def}), (\ref{eq:drho-lcit}), and (\ref{eq:hcit}) in the above expression, it is straightforward to show that
\be
{\mathcal R}_{jk}^{(2)}(t)=-\sum_{j'\neq k} \delta_{jk} {\mathcal W}_{k\rightarrow j'}^{(2)} (t)  + (1-\delta_{jk}) {\mathcal W}_{k\rightarrow j}^{(2)}(t) ,  \label{eq:Rjk-2}
\ee
where 
\ben 
{\mathcal W}_{j\rightarrow k}^{(2)}(t)&=&2{\rm Re}  \int_0^t d\tau {\rm Tr}_b \left\{ \hat H_{kj}^\dagger (t)\hat U_{0,k}(t-\tau) \hat H_{kj}(\tau) \right . \nonumber \\
&&\left . \hspace{0.8in} \times \hat U_{0,j}(\tau) \hat \rho_{b,j} \hat U_{0,j}^\dagger (t)\right\} . \label{eq:wjk-2}
\een
Employing Eq. (\ref{eq:Rjk-2}) in Eq. (\ref{eq:me-formal-1}) results in the form of Eq. (\ref{eq:me}) in the main text.  Equation (\ref{eq:wjk-2}) becomes equivalent to  Eq. (\ref{eq:kif-time-1}) for $\hat \rho_f=|\varphi_k\rangle\langle \varphi_k|\otimes \hat 1_b$,  $\hat \rho_j= |\varphi_j\rangle\langle \varphi_j|\hat \rho_{b,j}$, and $\hat \rho_{b,j}$ commutes with $\hat U_{0,j}(t)$.

\acknowledgments

SJJ acknowledges primary support from the US Department of Energy, Office of Sciences, Office of Basic Energy Sciences (DE-SC0026114).  SJJ also thanks the support of  the Korea Institute for Advanced Study (KIAS) through KIAS Scholar program, and \rc{Marshall D. Newton, Abraham Nitzan, and Eitan Geva for their valuable comments on the earlier version of this work.}   YMR acknowledges support from the National Research
Foundation of Korea (NRF) grant funded by the Korean government (MSIT) through SRC (No. RS-2020-NR049542).

 \vspace{.2in}

\noindent
{\bf  AUTHOR DECLARATIONS} \vspace{.1in}\\
{\bf Conflict of Interest} \vspace{.1in}\\
The authors have no conflicts to disclose. \vspace{.2in}\\
\noindent
{\bf  DATA AVAILABILITY} \vspace{.1in}\\
Most data that support the findings of this article are contained in this article.  Additional data are available from the corresponding author upon reasonable request.


\end{document}